\documentclass[AMA,Times1COL]{WileyNJDv5} %STIX1COL,STIX2COL,STIXSMALL
\usepackage{orcidlink}
\usepackage{hyperref}
\usepackage{makecell}
\usepackage{subcaption}
\usepackage{scalerel}
\usepackage{tikz}
\usepackage{amsmath}
\usepackage{amssymb}
\usetikzlibrary{svg.path}

\definecolor{orcidlogocol}{HTML}{A6CE39}
\tikzset{
  orcidlogo/.pic={
    \fill[orcidlogocol] svg{M256,128c0,70.7-57.3,128-128,128C57.3,256,0,198.7,0,128C0,57.3,57.3,0,128,0C198.7,0,256,57.3,256,128z};
    \fill[white] svg{M86.3,186.2H70.9V79.1h15.4v48.4V186.2z}
                 svg{M108.9,79.1h41.6c39.6,0,57,28.3,57,53.6c0,27.5-21.5,53.6-56.8,53.6h-41.8V79.1z M124.3,172.4h24.5c34.9,0,42.9-26.5,42.9-39.7c0-21.5-13.7-39.7-43.7-39.7h-23.7V172.4z}
                 svg{M88.7,56.8c0,5.5-4.5,10.1-10.1,10.1c-5.6,0-10.1-4.6-10.1-10.1c0-5.6,4.5-10.1,10.1-10.1C84.2,46.7,88.7,51.3,88.7,56.8z};
  }
}

\newcommand\orcidicon[1]{\href{https://orcid.org/#1}{\mbox{\scalerel*{
\begin{tikzpicture}[yscale=-1,transform shape]
\pic{orcidlogo};
\end{tikzpicture}
}{|}}}}

\newcommand{\phone}[1]{Phone: \texttt{#1}}

\articletype{Article Type}%

\received{Date Month Year}
\revised{Date Month Year}
\accepted{Date Month Year}
\journal{Journal}
\volume{00}
\copyyear{2025}
\startpage{1}

\begin{document}

\title{Negative Control Outcome Adjustment in Early-Phase Randomized Trials: Estimating Vaccine Effects on Immune Responses in HIV Exposed Uninfected Infants}

\author{Ethan Ashby\textsuperscript{1,2}\,\orcidlink{0009-0001-3713-4489},
Bo Zhang\textsuperscript{2}\,\orcidlink{0000-0002-4381-1124},
Genevieve G Fouda\textsuperscript{3}\,\orcidlink{0009-0005-8094-9332},
Youyi Fong\textsuperscript{2}\,\orcidlink{0000-0002-4230-6863},
Holly Janes\textsuperscript{2}\,\orcidlink{0000-0002-3237-984X}
}

\authormark{Ashby \textsc{et al.}}
\titlemark{NCO Adjustment in Early-Phase RCTs}

\address[1]{\orgdiv{Department of Biostatistics}, \orgname{University of Washington}, \orgaddress{\city{Seattle}, \state{WA}, \country{USA}}}

\address[2]{\orgdiv{Vaccine and Infectious Disease Division}, \orgname{Fred Hutchinson Cancer Research Center}, \orgaddress{\city{Seattle}, \state{WA}, \country{USA}}}

\address[3]{\orgdiv{Department of Pediatrics}, \orgname{Weill Cornell Medicine}, \orgaddress{\city{New York}, \state{NY}, \country{USA}}}

\corres{Corresponding author: Ethan Ashby, \email{eashby@uw.edu} \newline
\phone{(925) 586-4191}}

\presentaddress{Hans Rosling Center for Population Health, 3980 15th Avenue NE, Box 351617, Seattle, WA 98195-1617, USA}

%\fundingInfo{Text}
%\JELinfo{ejlje}

\abstract[Abstract]{Adjustment for prognostic baseline variables can reduce bias due to covariate imbalance and increase efficiency in randomized trials. While the use of covariate adjustment in late-phase trials is justified by favorable large-sample properties, it is seldom used in small, early-phase studies, due to uncertainty in which variables are prognostic and the potential for precision loss, type I error rate inflation, and undercoverage of confidence intervals. To address this problem, we consider adjustment for a valid negative control outcome (NCO), or an auxiliary post-randomization outcome believed completely unaffected by treatment but more highly correlated with the primary outcome than baseline covariates. We articulate the assumptions that permit adjustment for NCOs without producing post-randomization selection bias, and describe plausible data generating models where NCO adjustment can improve upon adjustment for baseline covariates alone. In numerical experiments, we illustrate performance and provide practical recommendations regarding model selection and finite-sample variance corrections. We apply our methods to the reanalysis of two early-phase vaccine trials in HIV exposed uninfected (HEU) infants, where we demonstrate that adjustment for auxiliary post-baseline immunological parameters can enhance precision of vaccine effect estimates relative to standard approaches that avoid adjustment or adjust for baseline covariates alone.}


\jnlcitation{\cname{%
\author{Ashby E, Zhang, B, Fouda, GG, Fong, Y, Janes, H}.
\ctitle{Negative Control Outcome Adjustment in Early-Phase Randomized Trials: Estimating Vaccine Effects on Immune Responses in HIV Exposed Uninfected Infants.} \cjournal{\it Journal.} \cvol{2025;00(00):X--XX}.}}

\maketitle

%\theshorttitle{\textbf{Short title}: NCO Adjustment in Early-Phase RCTs}

\renewcommand\thefootnote{}
\footnotetext{\textbf{Abbreviations:} HEU, HIV Exposed Uninfected; RCT, randomized controlled trial; NCO, negative control outcome; bAb, binding antibody.}

\renewcommand\thefootnote{\fnsymbol{footnote}}
\setcounter{footnote}{1}

\section{Background}\label{sec1}

\par Vertical transmission of HIV-1 remains a global challenge in efforts to slow the HIV pandemic. Despite over a decade of use of highly effective antiretrovirals (ARVs) to prevent vertical transmission, 160,000 infant infections still occur annually worldwide,\cite{UNAIDS_2020} with more than a third due to breastmilk transmission.\cite{miotti_hiv_1999, Bertolli_1996} Identifying a safe, effective HIV-1 vaccine that can be administered to infants after birth to prevent transmission via breastmilk would be a major step towards eliminating perinatal HIV. Further, ``germline targeting" HIV vaccines that induce broadly neutralizing antibodies (bnAbs) are under investigation, and recent work suggests that infants develop bnAbs more readily than adults.\cite{goswami_harnessing_2020, fouda_immunological_2020, simonich_hiv-1_2016} Comparing immune responses in infants and adults is informative for advancing both infant and adult HIV vaccine programs.
\par A key step towards identifying an effective HIV vaccine is evaluating its ability to induce cellular and humoral immune responses to the HIV-1 virus (hereafter referred to as a vaccine's immunogenicity). Evaluating vaccine immunogenicity in infants presents unique challenges. Unlike HIV-1 naïve adults, in whom humoral immune responses to HIV-1 under placebo can be assumed to be nearly zero,\cite{Follman_2006} HIV-exposed infants passively acquire HIV-specific antibodies from their mothers through the placenta. One approach to address this challenge is to measure antibody levels long enough after birth (roughly 12 months in most infants)\cite{Fouda_2015} such that maternally inherited antibodies have waned. However, this approach limits evaluation of the dynamics of vaccine-induced antibody responses which may be useful for informing the optimal number and/or timing of vaccine doses. Another approach is to compare antibody responses between infants randomly assigned to vaccine or placebo arms and to estimate the average vaccine effect on an immune marker of interest. However, small trial sizes, imbalanced randomization ratios used to maximize data on vaccine safety, and variability of antibody responses in both vaccine and placebo arms can yield imprecise vaccine effect estimates and low statistical power. More precise vaccine effect estimates could improve future vaccine evaluations in HIV-exposed infants and other populations with prior pathogen exposure.
\par One strategy to improve the precision of vaccine effect estimates in early-phase trials is to exploit methods for covariate adjustment. Covariate-adjusted estimators of treatment effects can achieve asymptotic precision gains, supporting their application to large, late-phase trials.\cite{tsiatis_covariate_2008, Ye_2022} Covariate adjustment may be especially beneficial in early-phase trials where sample sizes are constrained, chance covariate imbalance is more likely, and power to detect treatment effects may be low. However, covariate adjustment is seldom used in early-phase trials, since adjusting for weakly prognostic covariates can lead to precision loss, and adjustment for too many predictors can result in Type I error rate inflation and undercoverage of confidence intervals.\cite{Tackney_2023, kahan_risks_2014, van_lancker_use_2023} Current regulatory guidance recommends adjustment for a small number of covariates believed to be strongly associated with the outcome of interest.\cite{FDA} In our application to HIV-1 vaccine studies, pooled analyses of several trials in adults have struggled to identify reliable baseline predictors of HIV-1 vaccine-elicited immune responses. The most prognostic baseline predictors (sex, age, and body mass index) exhibit at-best modest correlations with HIV-1-specific immune responses,\cite{Huang_2017, huang_baseline_2022} and the applicability of these findings to infant studies remains unclear.
\par Auxiliary immunological measurements have the potential to more reliably predict vaccine-specific immune responses than commonly collected baseline variables. In the context of immune correlates analyses, Follmann proposed imputing missing vaccine-elicited immune responses for placebo recipients using immune responses to an unrelated ``baseline irrelevant vaccination".\cite{Follman_2006} Positive correlation between immune responses to HIV-1 vaccines and tetanus toxoid and hepatitis B vaccinations have been observed in adults.\cite{huang_non-hiv_2024} In populations with prior exposure to the pathogen of interest, baseline immune responses may be correlated with vaccine-elicited responses. Immune responses to herpes and influenza antigens at baseline were prognostic for immune responses at 6 weeks and 30 days after vaccination respectively.\cite{gilbert_hai_2019, gilbert_fold_2014} Vaccine studies in HIV-exposed uninfected (HEU) infants have the unique potential to collect a variety of auxiliary immunological measurements. Baseline HIV-1-specific immune responses in both the infant and mother can serve as proxies for the level of passively-inherited antibody at birth. Immune responses to ``off-target" HIV-1 antigens not included in the vaccine can describe the decay in maternal antibody over time. Immune responses to unrelated routine childhood vaccinations (e.g., tetanus, polio, haemophilus influenza B, hepatitis B, or measles) can provide proxy measurements of an infant's immune function, or the potential to mount immune responses to the HIV-1 vaccine.
\par Hence, in an early-phase vaccine study, an analyst may be tempted to adjust for a small set of predictive auxiliary immunological measurements instead of baseline covariates to improve precision of the estimated vaccine effect. However, adjusting for variables measured post-randomization in a randomized trial is typically discouraged due to the potential of inducing post-treatment selection bias in treatment effect estimates.\cite{Rosenbaum_1984} We argue that adjustment for post-baseline variables can be accomplished while avoiding selection bias if the variable is completely unaffected by treatment. Borrowing terminology from the causal inference literature, we refer to these post-baseline variables unaffected by treatment as negative control outcomes (NCOs).\cite{lipsitch_negative_2010, shi_selective_2020} We focus on early-phase trials where sample sizes are small, data on the control condition is constrained by unequal randomization ratios, and precision gains are most sought.
\par Our work offers the following contributions: (1) clear articulation of the assumptions enabling adjustment of post-baseline NCO, (2) demonstration of the asymptotic normality and semiparametric efficiency of the estimator that adjusts for a NCO, (3) causal graphical models describing instances where NCO adjustment can outperform baseline covariate adjustment, (4) a simulation study examining performance of NCO adjustment and providing guidance on model selection and finite-sample variance corrections, and (5) comparison of our method to existing methods in the reanalysis of two early-phase HIV vaccine trials in HEU infants.

\section{Methods}\label{sec2}

\subsection{Potential Outcomes Framework}

\par Consider a trial which randomly assigns a binary treatment $A$ and measures a continuous primary outcome $Y$. Our arguments can be extended to cases where $A$ takes $k$ distinct levels or where $Y$ is binary or count-based.\cite{zhang_improving_2008} In addition to the primary outcome, suppose that the study measures baseline covariates $X$ and an auxiliary outcome $N$. In our application, $A$ represents an HIV-1 vaccine, and $Y$ could be an antibody response to a HIV-1-specific antigen included in the vaccine construct. $X$ could include baseline variables like infant birth weight and sex, while $N$ could be an immune response to a routine childhood vaccine, an antibody response to a HIV-1-specific antigen not included in the vaccine, or a baseline immune response. Let $(Y(0), Y(1))$ and $(N(0), N(1))$ denote the potential outcomes for the primary and auxiliary outcomes under hypothetical placebo and vaccination.
\par With a continuous outcome and binary treatment, a typical causal estimand of interest is the population average treatment effect (ATE).
\begin{align*}
    \text{ATE} &= \mathbb{E}_{0}[Y(1) - Y(0)] 
\end{align*}
\par Suppose we adopt the following assumptions.

\vspace{4mm}

\textbf{Assumption 1}: Stable Unit Transform Value Assumption (SUTVA).\cite{Rubin_1980} SUTVA implies the following hold (i) no interference between participants, meaning that the potential outcomes of one participant does not depend on the treatment status of another, and (ii) no multiple versions of treatment. In some studies which measure infection or disease endpoints, the assumption of no interference between study participants is often questioned because an individual's infection outcome may depend on the vaccination statuses of their close contacts.\cite{HalloranLonghiniStruchiner_1999} However, our application focuses on immune response endpoints, which can safely be considered to be independent. No multiple versions of treatment is satisfied in double-blinded randomized trials with a sufficiently well-defined intervention. Under SUTVA, the observed data can be linked to the potential outcomes by the notion of consistency of potential outcomes; $Y = Y(A)$.

\vspace{4mm}

\textbf{Assumption 2}: Known treatment probability.\cite{Rosenbaum_1983}
\begin{align*}
    \pi = P(A=1) \text{ is known and lies between 0 and 1} 
\end{align*}
Assumption 2 implies that the treatment assignment is non-deterministic for all participants and is known by design. This is satisfied in the vast majority of randomized trials.

\vspace{4mm}

\textbf{Assumption 3}: Strong ignorability.\cite{Rubin_1978}
\begin{align*}
    A \perp (Y(0), Y(1), N(0), N(1), X)
\end{align*}
Randomization ensures that treatment is independent of potential outcomes, covariates, and even unmeasured characteristics. Blinding participants, caregivers, and personnel involved in data collection and analysis prevents latent sources of post-randomization bias (e.g., participant behaviors, outcome ascertainment by caregivers, participant/caregiver expectation of treatment benefit, etc.) from affecting participant outcomes. Herein, we focus on randomized, double-blinded trials where the potential for bias due to incomplete blinding is unlikely.

\par The above assumptions are considered satisfied in well-designed and conducted randomized trials and serve as the basis for identification of the population ATE from randomized trial data. The following assumptions are not required for identification but can be leveraged to improve the precision of inference.

\vspace{3mm}

\textbf{Assumption 4}: Negative control outcome (NCO).\cite{shi_selective_2020, Rosenbaum_1984}
\begin{align*}
    N = N(0) = N(1)
\end{align*}
Assumption 4 is strong and not typically invoked in randomized trials, because it implies zero treatment effect on the auxiliary outcome in every participant. Because $N$ is unaffected by treatment assignment, it can be effectively considered as a baseline covariate in subsequent developments.

\textbf{Assumption 5}: No effect of primary outcome $Y$ on NCO $N$. We use a pair of nonparametric structural equations models (NPSEMs) to formalize Assumption 5. Suppose potential outcomes are generated by a series of deterministic structural equations that depend on observed variables ($X, A$), possibly multivariate latent variables ($U_N, U_Y$), and random errors ($\epsilon_N \perp \epsilon_Y$).
\begin{equation}\label{eq:NPSEM1}
\begin{split}
    N &= f_N(X, U_N, \epsilon_N) \\
    Y(a) &= f_Y(X, A=a, N, U_Y, \epsilon_Y)
\end{split}
\end{equation}
Where $f_N$, $f_Y$ are unknown deterministic functions. We highlight three aspects of the NPSEMs. First, the NPSEMs are compatible with Assumption 4 because the structural equation for $N$ does not depend on the treatment variable $A$. Second, the structural equation for $N$ does not depend on the realization of the primary outcome $Y$. If the structural equation for $N$ depended on $Y$, Assumption 4 would be violated because $Y$ is affected by treatment. In the context of infant vaccine studies, $N$ may represent a baseline immune response or a proxy for maternally inherited antibodies, while $Y$ is a vaccine-elicited antibody response. In this case, $N$ is unlikely to be affected by $Y$ If $N$ is an immune response to an unrelated vaccine or an antigen not included in the vaccine, interference between antibody responses is possible, although evidence supports that infants and adults can mount immune responses to multiple distinct antigens simultaneously.\cite{itell_development_2018} Third, we allow the structural equation for $Y(a)$ to depend on the observed values of $N$. Suppose $N$ is a proxy for maternally inherited antibody. Previous research has shown that the presence of maternal inherited antibodies can ``blunt" vaccine-elicited immune responses.\cite{kandeil_immune_2020} However, because maternally inherited antibodies are unaffected by vaccination, $N$ may affect $Y$ without introducing bias. In Subsection \ref{subsec:NCOAdj}, we use causal directed acyclic graphs (DAGs) synonymous with the NPSEMs in Equation \ref{eq:NPSEM1} to motivate the potential benefits of adjusting for $N$ in a randomized experiment.

\subsection{Baseline Covariate Adjustment}

\par Consider a randomized trial with binary treatment $A$, baseline covariate $X$, and primary outcome $Y$. We assume the observed data are $n$ independent and identically distributed (i.i.d.) samples from some distribution $P_0$ in statistical model $M$.
\begin{align*}
    O_i = (A_i, X_i, Y_i) \overset{iid}{\sim} P_0 \in M
\end{align*}
A natural estimator for the population ATE is the plug-in estimator, or the difference in means between the two treatment groups.
\begin{equation}\label{eq:plug-in}
\begin{split}
    \hat{\psi}_{\text{plug-in}} &= \bar{Y}(1) - \bar{Y}(0) = \frac{1}{n} \sum_{i=1}^n \frac{A_i Y_{i}}{\hat{\pi}} - \frac{(1-A_i) Y_{i}}{(1-\hat{\pi})}
\end{split}
\end{equation}
\par Where $\bar{Y}(0)$ and $\bar{Y}(1)$ are the observed mean outcomes in the placebo and treatment group respectively, and $\hat{\pi}:=\frac{1}{n} \sum_{i=1}^n A_i$ is the observed treatment probability. Tsiatis and colleagues\cite{tsiatis_covariate_2008} demonstrated that any reasonable estimator of the population average treatment effect in a randomized trial measuring baseline covariates $X$ either belongs to or is asymptotically equivalent to a member of the following class of augmented inverse probability weighted (AIPW) estimators.
\begin{equation}\label{eq:augclass}
    \hat{\psi}_{\text{AIPW}} \in \left\{ \hat{\psi}_{\text{Plug-in}} - \frac{1}{n} \sum_{i=1}^n \left(A_i - \hat{\pi}\right) \left\{\frac{h_0(X_i)}{1-\hat{\pi}} + \frac{h_1(X_i)}{\hat{\pi}} \right\} : \text{Var}(h_a) < \infty \right\}
\end{equation}
\par Where $h_0$ and $h_1$ are arbitrary functions with finite variance. The AIPW estimators modify the plug-in estimator by subtracting a mean-zero augmentation term that exploits baseline covariates to improve precision. Tsiatis and colleagues\cite{tsiatis_covariate_2008} derived the smallest possible variance achievable by an estimator in the class in \ref{eq:augclass}. The \textit{semiparametric efficient estimator} achieves the minimum variance bound, and is given by $\hat{\psi}_{\text{AIPW}}$ with $h_a = \mathbb{E}[Y|A=a, X]$. The derivation of the variance bound and efficient estimator are based on arguments rooted in semiparametric theory, a review of these concepts and the derivation can be found in the Supplementary materials.
\par In practice, one must propose models for the unknown functions $h_a(X)$. In the case of a continuous outcome where the number of covariates is small, a common approach uses ordinary least squares (OLS) regressions for $\hat{h}_a(X)$. Models are fit separately in each arm or equivalently with full treatment-covariate interactions to ensure large-sample optimality.\cite{tsiatis_covariate_2008} The resulting estimator $\hat{\psi}_{\text{Cov-AIPW}}$ is consistent and asymptotically normal even if the models for $h_a(X)$ are misspecified, because the estimator belongs in the class of estimators in Equation \ref{eq:augclass}. Hence, $h_a(X)$ are often referred to as \textit{working models}, as correct specification is not required to attain consistency and normality. When OLS-working models are used, $\hat{\psi}_{\text{Cov-AIPW}}$ achieves guaranteed efficiency gain relative to plug-in estimators in large samples.\cite{Ye_2022} Alternatively, nonlinear regression models (e.g., GLMs or cross-fitted machine learning estimators) could be used as working models for adjustment, and can achieve guaranteed efficiency gain when paired with an OLS calibration procedure applied to the model predictions.\cite{Cohen2024-jp, Bannick2023-ge}
\par To obtain valid confidence intervals and hypothesis tests, a consistent estimator of the variance of $\hat{\psi}_{\text{Cov-AIPW}}$ is required. Motivated by our application to early-phase trials, we also desire a variance estimator that is robust to heteroskedasticity, misspecification of the working models, and exhibits good performance in finite samples. Tsiatis advocated for the use of sandwich standard errors with a degrees-of-freedom correction to account for the additional variability induced by model-fitting.\cite{tsiatis_covariate_2008} We consider the general variance estimator.\cite{MacKinnon_White_1985}
\begin{align*}
    \widehat{\text{Var}}(\hat{\psi}_{\text{Cov-AIPW}}) 
    &= (\textbf{C} \odot \textbf{R})^T \cdot (\textbf{C} \odot \textbf{R})
\end{align*}
Where $\odot$ denotes element-wise product, $\textbf{C} = (C_1, \ldots, C_n)^T $ is a vector of finite-sample correction factors, and $\textbf{R} \in \mathbb{R}^n$ are the usual components of the sandwich variance with entries
\begin{align*}
    R_i &= \left(\frac{A_i}{\sum_{j=1}^n A_j} - \frac{(1-A_i)}{\sum_{j=1}^n (1-A_j)}\right) Y_i - \frac{\hat{\psi}_{\text{AIPW}}}{n} - (A_i - \hat{\pi})\left(\frac{\hat{h}_0(X_i)}{\sum_{j=1}^n (1-A_j)} + \frac{\hat{h}_1(X_i)}{\sum_{j=1}^n A_j}\right) - (A_i - \hat{\pi}) \left(\frac{\bar{Y}(0)-\bar{h}_0}{\sum_{j=1}^n (1-A_j)} + \frac{\bar{Y}(1)-\bar{h}_1}{\sum_{j=1}^n A_j}\right)
\end{align*}
Where $\bar{h}_0 := \frac{1}{n} \sum_{i=1}^n \hat{h}_0(X_i)$, $\bar{h}_1 := \frac{1}{n} \sum_{i=1}^n \hat{h}_1(X_i)$. While variance corrections become negligible in large samples, they may be helpful in avoiding undercoverage of confidence intervals and type I error rate inflation in finite samples.\cite{Tackney_2023} To improve finite-sample performance, we explore the following correction factors.
\begin{enumerate}
    \item HC0-type: $C_i=1$, no correction.
    \item HC1-type: $C_i = \left(\frac{(n_0 - p_0 - 1)^{-1} + (n_1 - p_1 - 1)^{-1}}{(n_0 - 1)^{-1} + (n_1 - 1)^{-1}}\right)^{1/2}$ where $p_0$ and $p_1$ are the number of parameters fit in models $\hat{h}_0$ and $\hat{h}_1$ respectively, and $n_1, n_0$ are the number of treated and control participants respectively.\cite{tsiatis_covariate_2008}
    \item HC2-type: $C_i := \left(1/(1-H_{a, i, i})\right)^{1/2}$ where $H_{a, i, i}$ is the leverage of observation $i$, or the $i$-th diagonal entry of the ``hat matrix" for the OLS regression in arm $A=a$.\cite{MacKinnon_White_1985}
    \item HC3-type: $C_i := 1/(1-H_{a, i, i})$ where $H_{a, i, i}$ is defined identically above.\cite{MacKinnon_White_1985}
\end{enumerate}

\subsection{Negative Control Outcome Adjustment}\label{subsec:NCOAdj}

Covariate adjustment in randomized experiments almost exclusively focuses on pretreatment variables. Adjusting for a post-treatment variable introduces the possibility that the variable was affected by treatment, which can result in selection bias. However, fixating on pretreatment variables is sufficient but not necessary to avoid bias. The hallmark of acceptable adjustment variables is not that they were measured before the intervention, but that they were unaffected by the intervention.\cite{Rosenbaum_1984} We formalize this notion using the NCO Assumption (Assumption 4). NCOs have been used extensively for bias detection and elimination in non-randomized studies,\cite{lipsitch_negative_2010, shi_selective_2020} but we consider how NCOs can be used to augment inference in randomized trials.
\begin{figure}[h]
\centering
\includegraphics[width=0.6\textwidth]{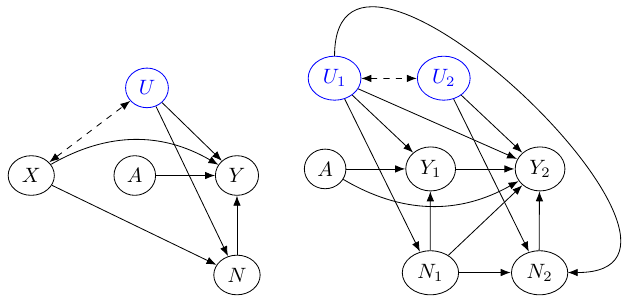}
\caption{Causal diagrams illustrating assumptions where NCOs are a proxy for a prognostic unmeasured precision variable $U$ (left) or prognostic unmeasured time-varying process $(U_1, U_2)$ (right). In the right panel, baseline covariates $X$ are omitted for simplicity.}
\label{fig:DAGs}
\end{figure}

We provide two examples of graphical causal models based on directed acyclic graphs (DAGs) where NCO adjustment may offer benefits relative to baseline covariate adjustment. First, we focus on the DAG in the left panel of Figure \ref{fig:DAGs}. The DAG corresponds to the NPSEMs in Equation \ref{eq:NPSEM1} when the unmeasured variables affecting the primary and negative control outcomes are identical, i.e., $U = U_Y = U_N$. In the literature on bias detection in observational studies using negative controls, the degree of overlap between $U_Y$ and $U_N$ is often referred to as the ``U-comparability" of $N$ and $Y$.\cite{shi_selective_2020} When $N$ and $Y$ are exactly U-comparable and the contributions of the errors to each outcome are small, $N$ is a reliable surrogate for the common unobserved variable $U$ affecting the primary outcome.\cite{Rosenbaum_1984} In reality, $N$ and $Y$ may only be approximately U-comparable (i.e., $U_N \approx U_Y$), but $N$ may still be a useful proxy for the effect of $U_Y$ on $Y$. In our application to infant HIV vaccine studies, an important missing variable ($U$) is the function of an infant's immune system at birth. While baseline covariates such as infant sex or birth weight will capture limited information on $U$, immune responses to routine childhood vaccinations ($N$) are unlikely to be affected by HIV-1 vaccination but may be relevant proxies for $U$. A second causal model for longitudinal data is described in the DAG in the right panel of Figure \ref{fig:DAGs}, and a description using NPSEMs can be found in the Supplementary materials. In the DAG, $N$ and $Y$ are exactly U-comparable and the tuple $(N_1, N_2)$ can be considered as a proxy for the time-varying unmeasured process ($U_1, U_2$). In infant vaccine studies, the decay of maternally inherited antibodies is a plausible example of a time-varying unmeasured process ($U_1, U_2, \ldots$) influencing primary outcomes. Longitudinal measures of immune responses to synthetic antigens or HIV-1 antigens not included in the vaccine ($N_1, N_2, \ldots$) may be useful surrogates for maternal antibody decay. Covariates measured at baseline may be particularly poorly equipped to explain how immune responses will evolve over time as maternal antibody decays.
\par We highlight a few reasons why adjustment for NCOs may be appealing from an analyst's perspective seeking to reduce finite-sample bias and improve precision of ATE estimates. First, an NCO measured during follow-up may be more predictive of the outcome of interest than baseline variables, especially when participant characteristics evolve post-baseline (e.g., due to developmental or maturation processes). If the NCO is a proxy for such changes, it may be a better choice of adjustment variable than baseline predictors. Second, the NCO may explain idiosyncratic, technical sources of variation in the primary outcome that cannot be explained by baseline variables. In our application to vaccine studies, binding antibody multiplex assays (BAMA) are often used to measure antibody responses to several antigens in parallel. A negative control antibody response measured using BAMA may explain variability in the primary antibody responses due to sample preparation and assay execution. Third, NCOs can simplify the problem of selecting which variables to adjust for. Randomized trials commonly collect many baseline covariates with questionable relevance to the outcome of interest. While adjustment for all predictors will not harm large-sample efficiency (see Proposition \ref{prop:NCOadj} below), it can lead to efficiency loss in finite-samples when variables exhibit low or moderate correlation with the primary outcome. Thus, adjusting for a small, valid set of predictive NCOs may be preferable to adjusting for a large set of baseline covariates in a small trial.

We present the following Proposition, which extends the asymptotic theory for covariate adjustment to include NCOs. A proof of Proposition \ref{prop:NCOadj} can be located in the Appendix.

\begin{proposition}\label{prop:NCOadj}
    Suppose Assumptions 1-5 hold. The following model-assisted estimator is consistent and asymptotically normal for the ATE for any choices of working models $\hat{h}_0, \hat{h}_1$ with finite variance.
    \begin{equation}\label{eq:ncocov-aug}
     \hat{\psi}_{\text{CovNCO-AIPW}} := \hat{\psi}_{\text{Plug-in}} - \frac{1}{n} \sum_{i=1}^n \left(A_i - \hat{\pi}\right) \left\{\frac{\hat{h}_0(X_i, N_i)}{1-\hat{\pi}} + \frac{\hat{h}_1(X_i, N_i)}{\hat{\pi}} \right\}
    \end{equation}
    If $\hat{h}_0$ and $\hat{h}_1$ are fit using OLS regression, the estimator will have guaranteed asymptotic efficiency gain over unadjusted estimators and estimators which adjust for $X$ alone. If $\hat{h}_0$ and $\hat{h}_0$ are correctly specified models for $\mathbb{E}[Y|X,N,A=0]$ and $\mathbb{E}[Y|X,N,A=0]$, $\hat{\psi}_{\text{CovNCO-AIPW}}$ will be semiparametric efficient, meaning it achieves the lowest asymptotic variance among all asymptotically linear estimators and data generating laws satisfying $A \perp (X,N)$.
\end{proposition}

Proposition \ref{prop:NCOadj} indicates that when Assumptions 4 and 5 hold, the AIPW estimator that adjusts for $(X,N)$ is the most efficient estimator of the ATE in large samples among trials that measure $(X,N)$. However, we can show that adjustment for an NCO alone can lead to semiparametric efficient inference in an augmented data model under additional assumptions detailed in the corollary below.

\begin{corollary}\label{cor:cor1}
    Consider a model $M^*$ encompassing data generating laws for a randomized trial measuring an oracle data unit that includes the unmeasured variable $U$: $\{X_i, N_i, U_i, A_i, Y_i\}_{i=1}^n \overset{i.i.d}{\sim} P_0 \sim M^*$. Suppose Assumptions 1-5 hold, but where Assumption 3 is modified to $\{Y(1),Y(0),N,X,U\} \perp A$. The following model-assisted estimator is consistent and asymptotically normal for the ATE for any choices of working models $\hat{h}_0, \hat{h}_1$ with finite variance.
    \begin{equation}\label{eq:nco-aug}
     \hat{\psi}_{\text{NCO-AIPW}} := \hat{\psi}_{\text{Plug-in}} - \frac{1}{n} \sum_{i=1}^n \left(A_i - \hat{\pi}\right) \left\{\frac{\hat{h}_0(N_i)}{1-\hat{\pi}} + \frac{\hat{h}_1(N_i)}{\hat{\pi}} \right\}
    \end{equation}
    When $\hat{h}_0, \hat{h}_1$ are OLS regression models fit separately within each treatment arm, then $\hat{\psi}_{\text{NCO-AIPW}}$ offers guaranteed efficiency gain relative to the plug-in estimator. 

    In addition to Assumptions 1-5, suppose we assume (i) $N$ and $Y$ are U-comparable, $U=U_Y=U_N$ in Equation \ref{eq:NPSEM1} of Assumption 5 and (ii) $N$ is a surrogate for the effect of $(X,U)$ on $Y$ within levels of $A$, implied by the following mean independence assumption.
    \begin{align*}
        E[Y|A, X,U,N] = E[Y|A,N]
    \end{align*}
     When $\hat{h}_0, \hat{h}_1$ are OLS regression models fit separately within each treatment arm, then $\hat{\psi}_{\text{NCO-AIPW}}$ exhibits guaranteed efficiency gain relative to the unadjusted and covariate-adjusted estimator using OLS-working models and no efficiency loss relative to the OLS-assisted estimator which adjusts for the oracle $\{X, U, N\}$. Furthermore, if $\hat{h}_0, \hat{h}_1$ are correctly specified models for $E[Y|N,A=0]$ and $E[Y|N,A=1]$, then $\hat{\psi}_{\text{NCO-AIPW}}$ is efficient in the oracle model, meaning it achieves the lowest asymptotic variance among asymptotically linear estimators across all data generating laws for the oracle data unit $\{A,X,U,N,Y\}$ satisfying $A \perp (X,U,N)$.
\end{corollary}

Corollary 1 highlights several important characteristics of the parsimonious estimator that only adjusts for the NCO. First, adjustment for a valid NCO will always lead to improved efficiency relative to an unadjusted estimator under no further assumptions. Second, if $N$ is a valid surrogate for both measured and unmeasured causes of $Y$, then adjusting for the NCO improves upon adjusting for baseline covariates. Moreover, it does not lose efficiency relative the oracle estimator that adjusts for all measured and unmeasured causes. Third, if the working models are specified correctly, adjustment for $N$ alone leads to the most efficient estimator among all estimators adjusting for $(X,U,N)$.

We acknowledge that the NCO assumption is a strong assumption not typically invoked when analyzing randomized trials. There exists a tension between choosing an auxiliary outcome that is sufficiently predictive of the primary outcome but that is not affected by the intervention. The most predictive auxiliary outcomes may be more likely to introduce bias. The plausibility of the NCO assumption may hinge on subject matter knowledge and evidence from prior experiments, which may be limited in early-phase trials. However, the NCO assumption may be easier to justify in vaccine studies in specific circumstances, owing to complete knowledge of the vaccine construct and the remarkable specificity of the adaptive immune response. Empirical evidence from late-phase efficacy trials of HPV\cite{Tota2020-ya} and COVID-19 vaccines\cite{Ashby_2024} have shown negligible vaccine effects in preventing infections caused by vaccine untargeted pathogens, supporting their use as NCOs. Previous HIV-1 vaccine studies in infants have shown that markers of maternal antibody\cite{Johnson2005-uh, McFarland2006-hd} and immune responses to routine childhood vaccinations\cite{itell_development_2018} are unaffected by HIV-1 vaccination. However, identifying valid NCOs may be more difficult in other circumstances. Many modern HIV-1 vaccines aim to elicit immune responses that confer broad protection against many HIV-1 strains. In such cases, immune responses to HIV-1 antigens not included in the vaccine may be risky choices of NCOs. Choosing an NCO immune marker sufficiently distinct from the vaccine's mechanism of action may be an easier task for vaccines with fewer components (e.g., mRNA/DNA, protein subunit, or conjugate vaccines) than more antigenically diverse constructs (e.g., live attenuated, inactivated, mosaic, or polyvalent). In some rare cases, vaccines can confer protection against pathogens unrelated to the target disease through a memory-like response developed in innate immune cells (e.g., BCG vaccination given to infants).\cite{Covian2019-de} Ultimately, the selection of NCOs should be based on scientific knowledge and available data from prior studies. Coordinated efforts to collect a set of candidate NCOs across several early-phase trials could aid efforts to screen for valid NCOs that are correlated with immune responses of interest.

If an analyst wants to check the NCO assumption prior to adjustment, we suggest employing a pretest. We restrict focus to tests conducted at the nominal Type I error rate ($\alpha=0.05$). Let $\tau^N = N(1) - N(0)$ refer to the individual treatment effect of vaccine on the candidate NCO. One can test the sharp causal null hypothesis implied by Assumption 4 that $\tau^N=0$ for all participants using a randomization test, which envisions the treatment assignment mechanism as the sole source of randomness in an experiment. Rejection of the sharp null indicates that the observed data are unlikely under the NCO assumption, and NCO adjustment should be avoided. However, failure to reject the sharp null cannot be interpreted as evidence that the NCO assumption is true, and small early-phase trials may possess low power to discriminate between valid and invalid NCOs. As an alternative, we consider an \textit{equivalence pretest} of the hypothesis that individual causal effects $\tau^N$ lie outside an analyst-defined ``equivalence window" centered at zero, $[-\epsilon, \epsilon]$ for some analyst-chosen threshold $\epsilon >0$. Formally, we define the null hypothesis as $H_0^{\text{equiv}}: \tau^N < -\epsilon \; \cup \; \epsilon < \tau^N$. Rejection of $H_0^{\text{equiv}}$ implies that individual treatment effects are not too large, while failure to reject $H_0^{\text{equiv}}$ implies that the analyst cannot rule out large individual treatment effects. Note that $H_0^{\text{equiv}}$ is a hypothesis of \textit{bounded} individual treatment effects, for which valid tests exist when the test statistic used for randomization inference satisfies certain regularity conditions.\cite{Caughey2024-jr} Unlike the sharp null, the equivalence null hypothesis assumes that $N$ is an \textit{invalid} NCO, endowing the equivalence test with skepticism towards candidate NCOs in small studies when small equivalence thresholds $\epsilon$ are chosen. To guide choices of $\epsilon$, we propose using fractions of measures of dispersion (e.g., range or variance) of the primary outcome pooled across treatment arms. In forthcoming numerical experiments, we will show the relative benefits and drawbacks of conducting pretests before NCO adjustment.

\subsection{Finite-sample inference}

\par The approaches developed above are based on a statistical model that assumes units are exchangeable draws from a hypothetical superpopulation, and the parameter of interest is a treatment effect defined in the superpopulation. In early-phase trials, the superpopulation sampling model may be implausible if the trial imposes strict eligibility criteria or if the participants represent a convenience sample. An analyst may wish to restrict focus to whether the intervention worked in the study participants themselves. We explore two different approaches to finite-population inference.

\subsubsection{Adjusted Estimators of the Sample Average Treatment Effect}

Adopting Neyman's randomization inference framework,\cite{Neyman_1921} the potential outcomes $(Y(1), Y(0))$ and predictors $(X,N)$ are assumed fixed, and treatment assignment mechanism is the sole source of randomness in the study. Our estimand of interest is the average treatment effect in the study participants themselves, or the sample average treatment effect (SATE).
\begin{equation}\label{eq:SATE}
    \text{SATE} = \frac{1}{n} \sum_{i=1}^n Y_i(1) - Y_i(0)
\end{equation}
One may be interested in estimating the SATE and testing the \textit{weak null hypothesis} that  $H_0^{\text{Weak}}: \text{SATE} = 0$.\cite{Neyman_1921} The SATE is not identified because only one potential outcome is observed per individual. However, the plug-in estimator given in Equation \ref{eq:plug-in} is unbiased for the SATE and a conservative variance estimator is $\frac{\hat{S}^2_{Y,A=1}}{n_1} + \frac{\hat{S}^2_{Y,A=0}}{n_0}$ where $\hat{S}^2_{Y,A=a}$ is the sample variance of the outcome in arm $a$.
Lin\cite{Lin_2013} proposed a family of linearly adjusted estimators, which we extend to include NCOs.
\begin{align*}
    \hat{\psi}_{\text{Lin}} &\in \left\{\frac{1}{n_1} \sum_{i=1}^n A_i (Y_i - \beta_1^T (X_i - \bar{X}, N_i - \bar{N})) - \frac{1}{n_0} \sum_{i=1}^n (1-A_i) (Y_i - \beta_0^T (X_i - \bar{X}, N_i - \bar{N}))\right\}
\end{align*}
Where $\bar{X}$ and $\bar{N}$ are the sample means of the baseline covariates and NCO in the study population respectively. The choice of $(\beta_0, \beta_1)$ that minimizes the variance is obtained by OLS regression of outcomes $Y$ on predictors $(X,N)$ separately within each treatment arm $A$, or equivalently with full treatment-covariate interactions. The resulting estimator is consistent, asymptotically normal, and yields efficiency gain relative to the plug-in estimator in large samples under bounded moment and asymptotically stable randomization probabilities.\cite{Lin_2013} Lin also advocated estimating the variance of $\hat{\psi}_{\text{Lin}}$ using the sandwich method, which is consistent, asymptotically conservative, and robust to heteroskedasticity and linear model misspecification.

There are some caveats to using Lin's estimator for estimating the SATE in finite samples. First, the adjusted estimator introduces finite-sample bias due to estimating $(\beta_0, \beta_1)$. The bias is absent when the parameters are known. However, Lin showed that the bias diminishes rapidly with increasing sample size and argued that the bias due to adjustment should be weighed against the bias from covariate imbalance. Second, in small samples and/or settings with high leverage points due to outlying covariate values, sandwich standard errors may be anti-conservative. Finite-sample variance corrections, as discussed above, can restore the coverage of confidence intervals (CIs) and Type I error rate control at the nominal level. Third, Lin's sandwich variance estimator cannot be used for superpopulation inference because it does not account for the additional uncertainty due to centering the predictors. In randomization inference however, covariate centering can be ignored because the mean of the predictors is assumed fixed. Variance estimators which account for covariate centering for superpopulation inference have been developed elsewhere.\cite{Ye_2022}

\subsubsection{Adjusted Randomization Tests}

We continue within Neyman's causal model, where all covariates and outcomes are fixed and the only source of randomness in the experiment is the treatment assignment mechanism. Rather than estimating a treatment effect, suppose interest lies in testing whether the intervention satisfies the sharp null hypothesis of a zero treatment effect in study participants.
\begin{align*}
    H_0^{\text{sharp}}: Y_i(1) - Y_i(0) = 0 \hspace{4mm} \text{for all i} \in \{1, \ldots, n\}
\end{align*}
Under $H_0^{\text{sharp}}$, both potential outcomes $\{Y_{i}(0), Y_{i}(1)\}$ are known for every participant. The exact distribution of any arbitrary test statistic $T: \{Y_i(0), Y_i(1)\}_{i=1}^n \rightarrow \mathbb{R}$ under $H_0^{\text{sharp}}$ can be obtained by computing the test statistic over all possible permutations of the treatment assignment vector $\textbf{A}$. If the total number of permutations of $\textbf{A}$ is prohibitively large, the null distribution can be approximated using Monte Carlo methods. The test statistic computed on the observed data can be compared to randomization distribution under $H_0^{\text{sharp}}$ to construct finite-sample exact tests of the sharp null hypothesis.

There are two main strategies for adjusting for predictors in randomization tests. We refer to the first approach as the pseudo-outcome approach. In this approach, an arbitrary algorithm $f$ regresses the outcomes $Y$ on non-treatment predictors ($X,N$) to produce residuals $r = Y - f(X,N)$. The residuals $\{r_i\}_{i=1}^n$ are then used as the basis for randomization inference in a covariate-free fashion. The pseudo-outcome approach can lead to improved power of the randomization test when the residuals are more stable and less dispersed than the outcomes.\cite{Rosenbaum_2002} Under the Neyman model, covariates, valid NCOs, and potential outcomes are presumed fixed. Therefore, functions of these variables, including the residuals computed from the regression fit, are also fixed.\cite{Rosenbaum_2002} The regression $f$ is not a stochastic model but is merely an algorithmic fit to fixed quantities and introduces no additional uncertainty to the experiment. The pseudo-outcome approach can be used with any regression algorithm -- linear regression,\cite{Tukey1993-qi} nonlinear regression,\cite{Gail1988-dh} or data-adaptive methods \cite{Raz1990-pp, Stephens2013-bt} with or without variable selection techniques.

We refer to the second strategy as the model-output approach. The model-output approach involves regressing the outcome $Y$ on predictors $(X,N)$ and treatment $A$, then using the coefficient on the treatment variable $A$ as the test statistic upon which to base randomization inference.\cite{Gail1988-dh, Zhao_2021} Compared to the pseudo-outcome approach which requires only one algorithmic fit, the model-output approach requires refitting a model to each dataset corresponding to permutations of the treatment assignment vector, and may therefore incur additional computational costs. In theory, any model could be used to summarize the treatment effect, but the canonical choice is the linear model. As described by Zhao and Ding, summarizing the treatment effect using the robust t-statistic based on Lin's estimator of the SATE with OLS models and full interactions between treatment and predictors offers a number of theoretical advantages.\cite{Zhao_2021} That is, we can base randomization inference on the following quantity
\begin{align*}
    T = \frac{\hat{\psi}_{\text{Lin}}}{\sqrt{\hat{S}_0^2/n_0 + \hat{S}_1^2/n_1}}
\end{align*}
Where $\hat{S}_a^2 := (n_a-1)^{-1} \sum_{i : A_i = a} (Y_i - \hat{Y}(a))^2$ is the sample variance in arm $A=a$. No additional finite-sample corrections are required to ensure that randomization inference using Lin's robust t-statistic provides an exact test for $H_0^{\text{sharp}}$ in finite samples. The procedure is also asymptotically valid for testing the weak null hypothesis of zero \textit{average} treatment effect in the sample ($H_0^{\text{weak}}: \text{SATE} = 0$), which significantly broadens the test's application beyond the sharp null. Third, the procedure is guaranteed to be more powerful than unadjusted randomization tests under all alternatives hypotheses even if the OLS models are misspecified.\cite{Zhao_2021} In subsequent numerical experiments, we will focus on the model-output strategy based on Lin's $t$-statistic, given its desirable theoretical properties.
    
\section{Numerical Experiments}\label{sec:numeric}

\par We simulated trials ranging from small to moderate in size: $n \in \{40, 60, 80, 100, 120\}$ and considered a randomization probability favoring the intervention ($\pi=0.8$). For simplicity, we assumed a binary treatment $A$ with constant additive treatment effect parameterized by $\beta=1$. Our simulated trials measured a single quantitative baseline covariate $X$, primary outcome $Y$, and an auxiliary outcome $N$ at a single point in time. We focus on estimating the population average treatment effect; results of numerical experiments for finite-sample inference can be found in the Appendix.

\par Each observation was generated as i.i.d. draws from the following model.
\begin{align*}
    (X,U) &\sim N\left( \mu = (0,0), \Sigma = \begin{pmatrix}
        1 & \rho_{(X,U)} \\
        \rho_{(X,U)} & 1
    \end{pmatrix}\right) \\
    g(N) &\sim \text{Norm}(\mu_N = \beta_0 + \beta_1^N X + \beta_2^N U , \sigma^2 = 1) \\
    Y(0) &\sim \text{Norm}(\mu_Y = \beta_0 + \beta_1^Y X + \beta_2^Y U , \sigma^2 = 1) \\
    Y(A) &= Y(0) + \beta A
\end{align*}

For the main simulation, the auxiliary outcome $N$ was assumed to be a valid NCO satisfying Assumptions 4 and 5. We consider cases where $g$ is either the identity function (Setting 1) or a logistic function, $g(x)=-\log(8/x - 1)$ (Setting 2). The logistic function implies a distribution saturated at both lower and upper detection limits, a scenario common in immunogenicity studies. We assumed $\rho_{(X,U)}=0$ implying $X \perp U$ and assumed $\beta_0=1$ was fixed. We also assumed that $\beta_1^N = \beta_1^Y$ and $\beta_2^N = \beta_2^Y$, implying that the measured and unmeasured variables affected $N$ and $Y(0)$ with equal magnitude. This is similar to additive equiconfounding assumptions invoked for bias adjustment using NCOs in observational studies.\cite{shi_selective_2020} Using properties of the multivariate normal distribution, we express the remaining data-generating parameters, $\beta_1$ and $\beta_2$, in terms of more interpretable quantities (details can be found in the Supplementary Materials).
\begin{align*}
    \beta_1 &= \sqrt{\frac{\rho_{(Y,X)}^2 (\beta_2^2 + 1)}{1-(\rho_{(Y,X)})^2}} \hspace{5mm}  \beta_2 = \sqrt{\frac{\rho_{(Y,N|X)}}{1-\rho_{(Y,N|X)}}} 
\end{align*}
Where $\rho_{(Y,X)}$ describes the correlation between $Y$ and $X$ and $\rho_{(Y,N|X)}$ captures the correlation between the NCO and placebo outcome over and above that explained by the measured covariate. The baseline covariate was weakly predictive ($\rho_{(Y,X)}=0.3$) and we varied $\rho_{(Y,N|X)} \in \{0, 0.3, 0.5, 0.8\}$ encompassing cases where the $N$ was not predictive, weakly predictive, moderately predictive, and strongly predictive of $Y(0)$. 

\par We compared the performance of the following estimators: (a) Plug-in: $\hat{\psi}_{\text{plug-in}}$; (b) Covariate-adjusted: $\hat{\psi}_{\text{AIPW}}$ with $\hat{h}_a(X)$ fit using OLS regression with covariate-treatment interactions; (c) NCO-adjusted: $\hat{\psi}_{\text{AIPW}}$ with $\hat{h}_a(N)$ fit using OLS regression with predictor-treatment interactions; (d) Quantile-NCO-adjusted: $\hat{\psi}_{\text{AIPW}}$ with $\hat{h}_a(N^*)$ fit using OLS regression with predictor-treatment interactions after the empirical quantile transform, $N^* = \hat{F}_{N}(N)$, was applied to the NCO; and (e) Fully adjusted: $\hat{\psi}_{\text{AIPW}}$ with $\hat{h}_a(X,N)$ fit using OLS regression with predictor-treatment interactions. Estimators were compared with respect to (i) absolute finite-sample bias as a function of sample size $n$ and $\rho_{Y,N|X}$ relative to the plug-in estimator; (ii) average confidence intervals coverage over varying sample sizes $n$ and finite sample corrections $C$; (iii) relative efficiency defined as the ratio of each estimator's estimated variance compared to the plug-in estimator; and (iv) Power and Type I error of Wald tests of the null hypothesis that the population average treatment effect was zero under varying effect sizes and $\rho_{Y,N|X}$. We summarized simulation results over 1000 simulation replicates per condition. 

\par
\begin{figure}[h]
    \centering
    \includegraphics[width=\textwidth]{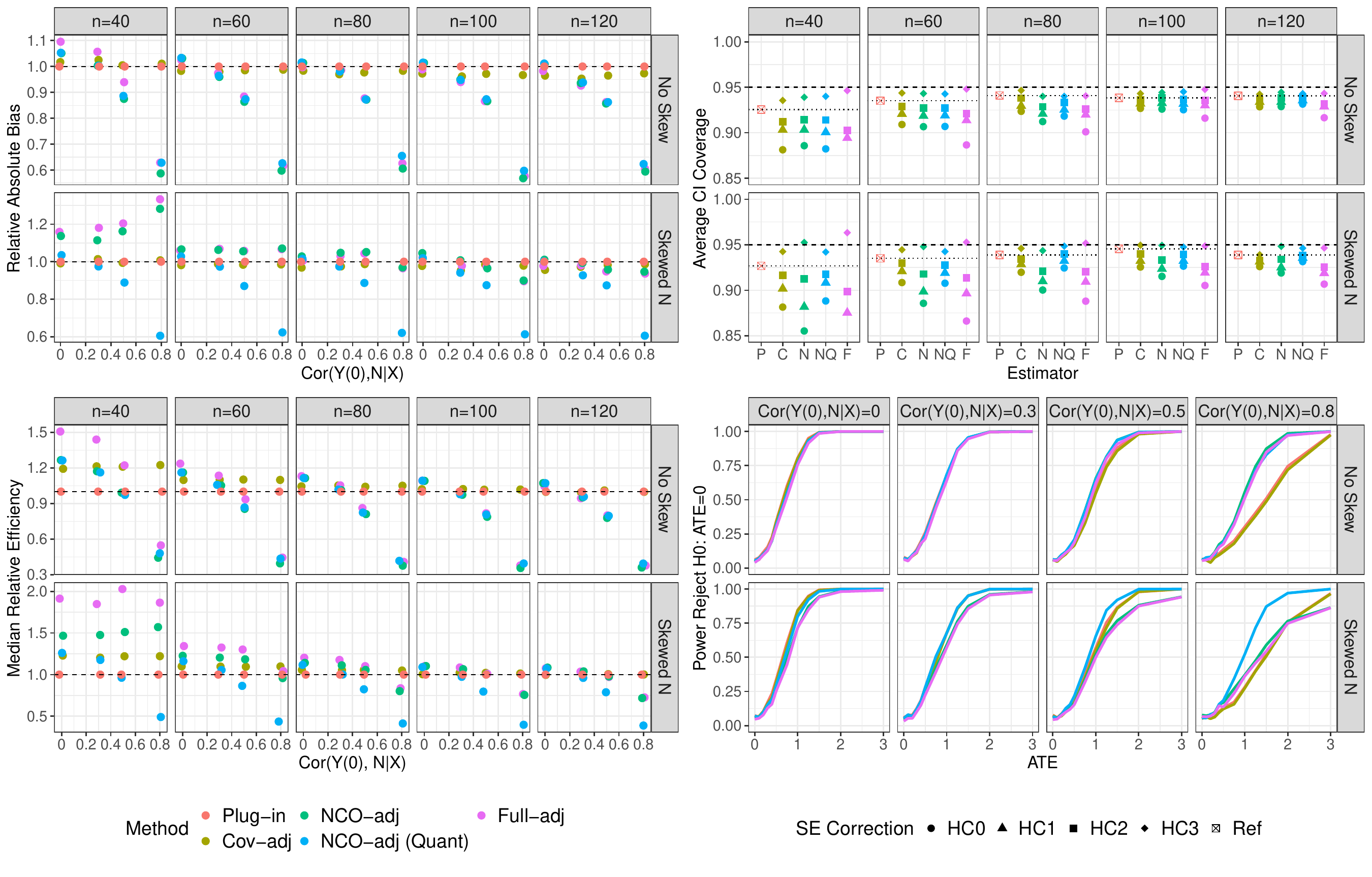}
    \caption{Top left: relative absolute bias of different estimators as a function of sample size, data generating mechanism, and predictiveness of NCO. Top right: average coverage of 95\% confidence intervals for different estimators with different finite-sample variance corrections as a function of sample size and data generating mechanism. Bottom left: median relative efficiency of different estimators as a function of sample size, data generating mechanism, and predictiveness of NCO. Bottom right: power curves of Wald tests of $H_0:\text{ATE}=0$ as a function of NCO predictiveness and the data generating mechanism. Sample size $n=60$ was fixed.}
    \label{fig:sim1}
\end{figure}

 The results of the numerical experiments are shown in Figure \ref{fig:sim1}. In the left panels of Figure \ref{fig:sim1}, NCO-adjusted estimators incurred greater finite-sample bias and lower precision in small trials when the NCO was skewed and not prognostic. Adjustment for both covariates and weakly prognostic NCOs generated additional bias and precision loss relative to simpler adjustment approaches. However, NCO-adjusted estimators achieved reductions in finite-sample bias and precision enhancement when sample sizes were moderate and the NCO was moderately correlated with the primary outcome. Adjustment for a quantile-transformed NCOs led to the largest finite-sample bias reduction and largest precision gain when the NCO was skewed. When the NCO was not skewed, adjustment for a quantile-transformed NCO still achieved meaningful improvements in performance. In the upper right panel of Figure \ref{fig:sim1}, the HC3 correction produced coverage estimates that matched or exceeded that of the plug-in estimator across all estimators, sample sizes, and data-generating mechanisms, supporting its use in early-phase trials. Lastly, in the bottom right panel, we observed that NCO adjustment can lead to improvements in power to test $H_0: \text{ATE}=0$ for $n=60$ even when the NCO was weakly or moderately predictive of the outcome of interest. Adjusting for a quantile-transformed NCO achieved the highest power when the NCO distribution was skewed and achieved power rivaling adjustment for an untransformed NCO when the NCO distribution was not skewed. Results for estimating the SATE and for randomization inference were very similar and can be located in the Supplementary materials.

\par We also conducted numerical experiments examining the performance of NCO adjustment when the NCO assumption (Assumption 4) was violated. Full details and results can be found in the Supplementary materials. We considered pairing the NCO-adjusted estimator with either a pretest of the sharp null hypothesis or an equivalence pretest with different equivalence thresholds $\epsilon > 0$. All pretests were conducted at level $\alpha=0.05$. If the sharp null hypothesis was rejected, the plug-in estimator would be returned, otherwise the NCO-adjusted estimator was returned. If the equivalence null hypothesis was rejected, the NCO-adjusted estimator was returned and otherwise, the plug-in estimator was returned. As anticipated, NCO adjustment resulted in bias when the treatment affected the candidate NCO. Checking the NCO assumption using a pretest partially mitigated the bias. A pretest of the sharp null protected against bias and poor coverage, especially when the NCO exhibited low correlations with the primary outcome. Equivalence pretests protected against violations of the NCO assumption outside the equivalence window $[-\epsilon, \epsilon]$. When NCO violations occurred within the equivalence window, bias occurred particularly for moderate size trials. When the NCO assumption was correct, pretests dampened the benefits of NCO adjustment. Pairing NCO adjustment with a pretest of the sharp null or an equivalence pretest (using a wide equivalence threshold) resulted in intermediate efficiency gains between the plug-in estimator and the NCO-adjusted estimator.

\par The implications of our simulation study for early-phase randomized trials are (1) NCO adjustment can lead to improved precision and power in small samples when the NCO is valid and is predictive of the primary outcome, (2) in the presence of skewed predictors subject to detection limits, adjustment for a quantile transformed NCO can mitigate finite-sample bias and variance inflation, (3) applying the HC3 correction to sandwich standard errors provided robust inference in finite-sample settings, and (4) despite the large-sample optimality of fully-adjusted estimators (as described in Proposition \ref{prop:NCOadj}), more parsimonious adjustment strategies offer better performance than fully-adjusted estimators in finite samples.

\section{Application to two early-phase vaccine trials in HIV-Exposed Uninfected Infant}

\par We apply our proposed methodology to two early-phase vaccine trials involving HIV exposed uninfected (HEU) infants. 

\subsection{HPTN 027}

\par
\begin{figure}[h]
    \centering
    \includegraphics[width=0.75\textwidth]{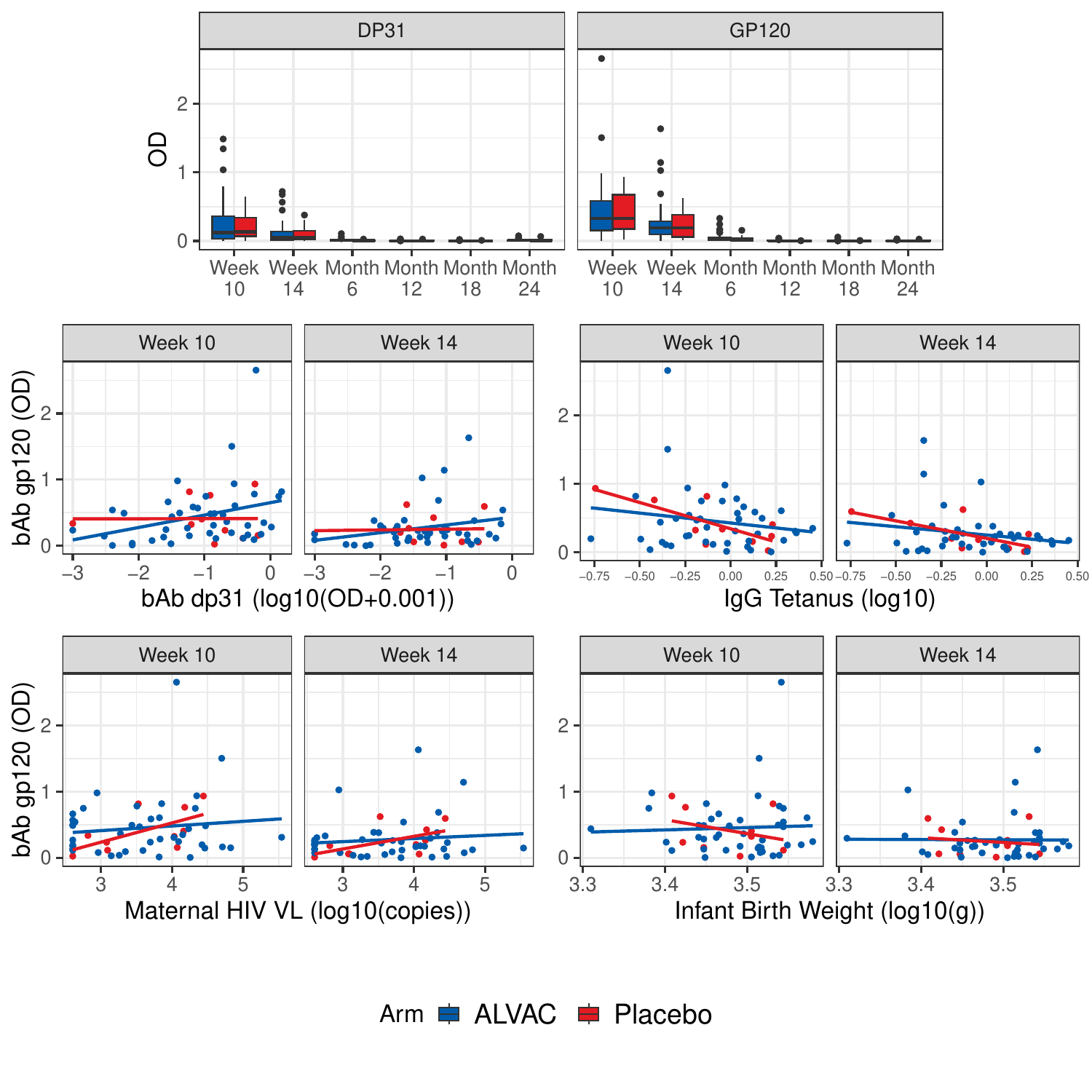}
    \caption{Top: boxplots of binding antibodies against synthetic antigen DP31 and target antigen GP120. Below: scatterplots and linear regression models stratified by treatment arm of candidate NCOs (middle rows) and baseline covariates (bottom rows).}
    \label{fig:htpn027}
\end{figure}

 The HIV Prevention Trials Network (HPTN) 027 was a randomized, double blind, placebo-controlled Phase I safety and immunogenicity study evaluating ALVAC-HIV vCP1521 (ALVAC) versus placebo in 60 HEU infants at Mulago National Referral Hospital in Kampala, Uganda.\cite{kaleebu_immunogenicity_2014} Notably, HPTN 027 was the first perinatal HIV vaccine trial conducted in Africa and evaluated the ALVAC vaccine used in the Thai RV144 trial, which demonstrated 30\% protection against HIV-1 acquisition in adults.\cite{rerks-ngarm_vaccination_2009} Infants were randomized to ALVAC (n=48) and saline placebo (n=12) and were immunized within 3 days of birth and at 4, 8, and 12 weeks. Several binding antibody measurements were recorded between 10 weeks and 24 months after birth. Our target of inference was average effect of ALVAC vaccination on binding antibodies (bAb) against gp120. Binding antibodies to DP31, a synthetic peptide not included in the vaccine, were also collected in parallel as a proxy for maternal antibody, which were mostly undetectable 6 months after birth. Infants received standard childhood vaccinations including the combination diphtheria tetanus pertussis (DTP) vaccine at 6, 10 and 14 weeks after birth. Antibodies to DTP vaccination were assessed once 6 months after birth. Immune responses were summarized and analyzed as background blank subtracted optical density (OD) values. We focused on estimating vaccine effects on gp120 antibodies at the Week 10 and 14 visits, which corresponded to assumed peak responses 2 weeks after Doses 3 and 4 of ALVAC respectively. Boxplots of immune responses to DP31 and GP120 over time and  scatterplots describing correlations between predictors of the antibody response to gp120 are shown in Figure \ref{fig:htpn027}, 
\par We compared the following ATE estimators: (i) plug-in, (ii) covariate-adjusted, (iii) NCO-adjusted, and (iv) a fully-adjusted estimator which adjusted for covariates and NCOs. All estimators used OLS working models fit separately in each treatment arm and HC3 variance corrections. We adjusted for two continuous baseline covariates: infant mass at birth and maternal HIV RNA PCR during the third trimester of pregnancy. We considered adjustment for two NCOs: the level of synthetic DP31 antibody at the same study visit and the antibody response to tetanus vaccination measured six months after birth. Pretests were not performed, since we believed HIV-1 vaccination would have zero effect on immune responses to synthetic or tetanus antigens. We applied log10 transforms to infant birth weight, maternal HIV RNA, and tetanus vaccination response. We applied a log10 transform to DP31 after adding 0.001 to all entries to avoid issues with zero DP31 responses.

\begin{table}[h]
\centering
\begin{tabular}{|c|c|c|c|c|c|c|}
\hline
Antigen & Contrast & Estimator & \makecell{Point Estimate\\(95\% CI) (OD)} & Variance & \makecell{Relative\\Efficiency} & Wald p-val\\
\hline
\multirow{6}{*}{\makecell{bAb gp120\\(Week 10)}} & \multirow{6}{*}{ALVAC v Placebo} & Plug-in & 0.042 (-0.201,0.285) & 0.0154 & 1.000 & 0.367\\
 &  & Covariate-adjusted & 0.046 (-0.199,0.291) & 0.0156 & 1.010 & 0.359\\
 &  & NCO-adjusted (DP31) & 0.045 (-0.194,0.284) & 0.0149 & 0.968 & 0.362\\
 &  & \textbf{NCO-adjusted (Tetanus)} & \textbf{0.043 (-0.156,0.242)} & \textbf{0.0103} & \textbf{0.673} & \textbf{0.334}\\
 &  & NCO-adjusted (DP31+Tetanus) & 0.044 (-0.167,0.254) & 0.0115 & 0.749 & 0.344\\
 & & Covariate+NCO-adjusted & 0.043 (-0.22,0.306) & 0.0180 & 1.170 & 0.377\\[0.5ex]
 \hline
\multirow{6}{*}{\makecell{bAb gp120\\(Week 14)}} & \multirow{6}{*}{ALVAC v Placebo} & Plug-in & 0.025 (-0.146,0.196) & 0.0076 & 1.000 & 0.389\\
 &  & Covariate-adjusted & 0.022 (-0.151,0.195) & 0.0078 & 1.020 & 0.385\\
 &  & NCO-adjusted (DP31) & 0.026 (-0.141,0.193) & 0.0073 & 0.952 & 0.383\\
 & & \textbf{NCO-adjusted (Tetanus)} & \textbf{0.025 (-0.116,0.166)} & \textbf{0.0052} & \textbf{0.679} & \textbf{0.364}\\
 &  & NCO-adjusted (DP31+Tetanus) & 0.025 (-0.119,0.169) & 0.0054 & 0.706 & 0.392\\
 &  & Covariate+NCO-adjusted & 0.02 (-0.151,0.191) & 0.0076 & 0.998 & 0.388\\
\hline
\end{tabular}
\caption{Estimated effect of HIV-1 ALVAC vaccine on IgG binding antibodies to vaccine-matched gp120 antigen in HPTN 027. Vaccine effects are estimated at Weeks 10 and 14, 2 weeks after vaccine doses 3 and 4. At each time point, six different estimates are provided corresponding to different predictor adjustment strategies. Vaccine effect is defined as the population ATE on the additive OD scale.}
\label{tab:hptn027}
\end{table}

Results for the data analysis are shown in Table \ref{tab:hptn027}. All estimators indicated that ALVAC vaccination was associated with an insignificantly higher average level of binding antibody to gp120 at Weeks 10 and 12. All point estimates were very similar, suggesting minimal bias due to adjustment for invalid or skewed NCOs. However, estimators that adjusted for immune response to tetanus vaccination achieved variance reductions exceeding 30\% relative to the plug-in estimator. Adjustment for antibodies to DP31 yielded more modest precision gains of 4 to 5\%. Adjustment for baseline covariates led to a loss in precision relative to the naive estimator. While large-sample theory supported the optimality of fully adjusted estimators, the estimator which adjusted for all predictors yielded lower precision, highlighting the penalty paid for adjusting for too many predictors in small samples.

\subsection{PACTG 230}

\par The Pediatric AIDS Clinical Trials Group (PACTG) 230 trial\cite{McFarland_2001, Fouda_2015} enrolled HEU infants in the United States to determine the safety and immunogenicity of two recombinant gp120 subunit protein vaccines: gp120-MN adsorbed into alum adjuvant (VaxGen) and rgp120-SF2 with MF59 adjuvant (Chiron). Infants were randomized to four injections of VaxGen rgp120 with alum (n=49), Chiron rgp120 with MF59 (n=48), and either placebo or adjuvant alone (n=19) between 0 and 20 weeks of age. For the purposes of our analysis, we grouped together all dose groups and dosing regimens within each platform. HIV Env-specific IgG were measured against HIV Env antigens at week 0, week 10, week 24, week 52, week 76, and week 104 after birth by BAMA. All outcomes were measured as blank-subtracted Mean Fluroescence Intensity (MFI) values. We focused on the week 24 immunogenicity visit corresponding to 4 weeks after the final dose of vaccine and focused on binding antibody IgG against the MN-gp120 antigen. The outcome was log10(MFI+1) transformed. We estimated the population ATE of vaccine on the primary outcome using the following estimators of the vaccine effect: (i) plug-in, (ii) covariate-adjusted, (iii) NCO-adjusted, and (iv) ``fully-adjusted" estimator which adjusted for covariates and NCOs.  We adjusted for a single baseline covariate: the baseline bAb level to gp120. We adjusted for a single negative control outcome: the contemporaneous level of IgG to gp41, an HIV-1 specific antigen not targeted by the Chiron or VaxGen vaccines. We employed an empirical quantile transform to baseline MN-gp120 IgG and to Week 24 gp41 IgG. All estimators used OLS working models fit separately in each treatment arm and HC3 variance corrections.

\begin{figure}
    \centering
    \includegraphics[width=0.8\textwidth]{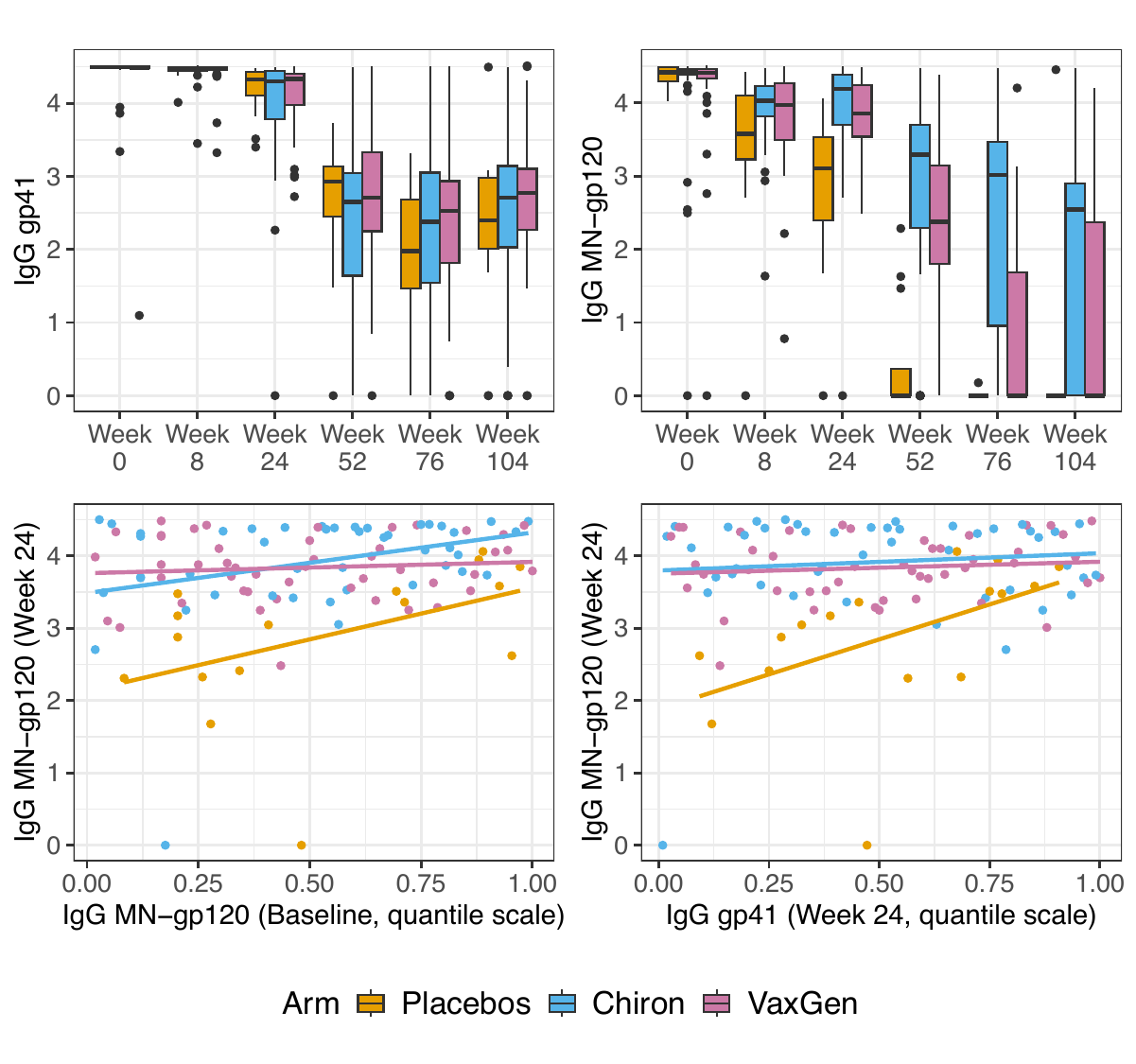}
    \caption{Above: boxplots of bAb against vaccine-untargeted HIV-1 antigen gp41 and vaccine targeted HIV-1 antigen MN gp120. Below: association between predictors and bAb to gp120 four weeks after Dose 4 of vaccine/placebo. OLS regressions within each arm shown.}
    \label{fig:fig2}
\end{figure}

Boxplots of the binding antibody responses to the primary antigen (MN-gp120) and negative control antigen (gp41) are shown in the left panel of Figure \ref{fig:fig2}. At baseline, most of the outcomes were at an upper limit/saturation point of the assay. Chiron and VaxGen vaccines generated higher bAb responses to gp120 relative to placebo. The gp41 response appeared very similar between the dose groups, supporting its use as a NCO. Furthermore, when outcomes were plotted against predictors, we observed that bAb to gp120 was positively correlated with the quantile of baseline bAb and quantile of Week 24 bAb. The positive association was most evident in the placebo arm. 

Despite the fact that gp41 was not included in either vaccine construct, we wanted to empirically test the NCO assumption prior to adjusting for the NCO. We performed a pretest of the NCO assumption $H_0^{\text{sharp}}: N(1)=N(0)$ on the bAb gp41 responses four weeks after the third dose using adjusted randomization tests. We adjusted for the baseline bAb response to gp120 and used the robust-t statistic based on Lin's adjusted estimator as the test statistic. We simulated the null distribution using 1000 random permutations of the treatment assignment vector. The pretest failed to reject the NCO assumption, as all randomization inference p-values were $>0.85$.

\begin{table}[h]
\centering
\begin{tabular}{|c|c|c|c|c|c|c|}
\hline
Antigen & Contrast & Estimator & \makecell{Point Estimate\\(95\% CI) (OD)} & Variance & \makecell{Relative\\Efficiency} & Wald p-val\\
\hline
\multirow{12}{*}{\makecell{bAb GP120\\(4 weeks post Dose 3)}} & \multirow{4}{*}{Chiron v Placebo} & Plug-in & 1.03 (0.48,1.57) & 0.0779 & 1.000 & 1.16e-04\\
 & & Covariate-adjusted & 1.02 (0.51,1.53) & 0.0688 & 0.883 & 5.12e-05 \\
 & & NCO-adjusted & 1.06 (0.55,1.57) & 0.0676 & 0.868 & 2.19e-05 \\
 & & \textbf{Covariate+NCO-adjusted} & \textbf{1.05 (0.55,1.55)} & \textbf{0.0657}  & \textbf{0.843} & \textbf{2.09e-05} \\
 & \multirow{4}{*}{VaxGen v Placebo} & Plug-in & 0.95 (0.43,1.47) & 0.0698 & 1.000 & 1.66e-04 \\
 & & Covariate-adjusted & 0.97 (0.49,1.46) & 0.0617 & 0.884 & 4.54e-05 \\
 & & \textbf{NCO-adjusted} & \textbf{0.98 (0.51,1.45)} & \textbf{0.0575} & \textbf{0.824} & \textbf{2.35e-05} \\
 & & Covariate+NCO-adjusted & 0.99 (0.52,1.46) & 0.0579  & 0.830 & 2.04e-05 \\
& \multirow{4}{*}{Chiron v VaxGen} & Plug-in & 0.08 (-0.17,0.33) & 0.0166 & 1.000 & 2.68e-01\\
& & Covariate-adjusted & 0.06 (-0.19,0.31) & 0.0163 & 0.976 & 3.14e-01 \\
& & NCO-adjusted & 0.08 (-0.18,0.34) & 0.0172 & 1.030 & 2.70e-01 \\
& & Covariate+NCO-adjusted & 0.06 (-0.19,0.32) & 0.0168  & 1.010 & 3.15e-01  \\
\hline
\end{tabular}
\caption{Estimated effect of gp120 vaccines on IgG binding antibodies to MN-gp120 antigen in PACTG230. Vaccine effects are estimated at Week 24, four weeks after final dose of study vaccines. Three contrasts are given, comparing Chiron and VaxGen vaccines to placebo and the two experimental vaccines head-to-head. For each contrast, four different estimates are provided corresponding to different adjustment strategies. Vaccine effect is defined as the population ATE on the log10(MFI+1) scale.}
\label{tab:pactg230}
\end{table}

Results of the data analysis are shown in Table \ref{tab:pactg230}. When comparing the Chiron and VaxGen vaccines to placebo, adjustment for baseline gp120 bAb led to efficiency improvements of approximately 12\% relative to the plug-in. Adjusting for Week 24 bAb to gp41 led to meaningful improvements in efficiency of approximately 13 to 18\%. Adjusting for both predictors led to improvements in precision between 16 to 17\%. All estimators had very similar point estimates, alleviating concern of bias due to skewed predictor distributions. However, adjustment for the predictors resulted in no clear improvements in efficiency when comparing the vaccines head-to-head.

\section{Discussion}

We highlight that adjustment for post-baseline negative control outcomes is an useful strategy to sharpen inference on treatment effects in early-phase randomized trials. NCO adjustment is particularly suitable for early-phase trials and secondary analyses, where resources are limited and a primary goal is to improve precision and power. From an analyst perspective, negative control outcomes may offer practical benefits when planning a strategy for adjustment, due to higher correlations with the primary outcome, simplification of the variable selection problem, and ability to capture hard-to-measure characteristics, unfolding developmental processes, or technical sources of variation that cannot be explained by baseline covariates. While large-sample theory supports the optimality of ``fully-adjusted" estimators which adjust for all available covariates and NCOs, simulations and application to two early-phase HIV-1 vaccine studies demonstrate that parsimonious adjustment for a limited set of NCOs can improve upon the performance of unadjusted, baseline covariate-adjusted estimators, and fully adjusted estimators in finite samples.

The main drawback of NCO adjustment is the possibility of introducing post-treatment selection bias. NCO-adjustment should only be considered when a candidate NCO is highly unlikely to be affected by the intervention based on subject-matter knowledge or prior experimentation. However, such information may be limited in early-phase trials. We argue that the NCO assumption can be justified in some cases, particularly in vaccine trials, due to the remarkable specificity of the adaptive immune response and complete analyst knowledge of the vaccine construct. If an analyst wishes to check their assumption, we propose pretests of either a sharp null hypothesis or an equivalence hypothesis to guard against violations of the assumption. The sharp null pretest can be used to identify NCO assumption violations in trials that are reasonably well-powered. If the trial is underpowered, an equivalence pretest with an appropriately chosen threshold may be preferred. In the Supplementary materials, we offer a simple sensitivity analysis based on linear structural mean models for the potential outcomes $Y(a)$ (which assume that the average treatment effect of $A$ on $Y$ is homogeneous in $N$). Sensitivity of experimental results to violations of the NCO assumption can be obtained by varying the unknown sensitivity parameter, the average treatment effect of $A$ on $N$, over a plausible grid of values. In future vaccine studies, coordinated data collection on a panel of candidate NCOs across multiple studies could help screen for NCOs that are unaffected by vaccination but are prognostic for immune responses of interest.

We primarily focus on treatment effect estimators which rely on linear working models with full predictor by treatment interactions fit using ordinary least squares, owing to their favorable theoretical properties. Recent investigations have proposed flexible, data-adaptive approaches for covariate adjustment.\cite{Bannick2023-ge, williams_optimising_2022, van_lancker_automated_2024} However, the application of such methods in early-phase trials is yet unclear. Supplying flexible machine learning estimators as working models will require sample-splitting to ensure Type I error rate control,\cite{Bannick2023-ge, van_lancker_automated_2024} which may not be feasible when trial sizes are small and randomization probabilities are imbalanced. The application of flexible models pretrained on historical data or variable selection algorithms such as stepwise selection or LASSO warrant further investigation in small studies. While randomization inference offers an opportunity to leverage data-adaptive methods for efficiency gain, it faces challenges, such as the questionable scientific importance of the sharp null. Randomization tests also do not immediately lend themselves to interval estimators of an interpretable treatment effect parameter. Recent work has used randomization test inversion to develop confidence intervals for quantiles of individual treatment effects,\cite{Caughey2024-jr} which may be particularly relevant to vaccine studies.\cite{chen_role_2024} Whether these approaches can be extended to incorporate adjustment for covariates and NCOs, perhaps even using flexible data-adaptive methods, is a possible direction for future work.

In the causal inference literature, ideal NCOs can be conceptualized as error-prone proxies of the treatment-free potential outcome. We distinguish between negative controls as either error-prone proxies of the treated or treatment-free potential outcome, and discuss their relative advantages. In HPTN027, antibodies to the synthetic antigen DP31 are a proxy for the level of maternal antibody and hence, the hypothetical HIV-1-specific antibody response under placebo (i.e., the treatment-free potential outcome). Adjustment for antibodies to DP31 did not reduce variance of the ATE, likely due to the lack of variability in gp120 responses in the trial's placebo arm. In contrast, antibody responses to tetanus vaccination can be considered as a proxy for an infant's hypothetical immune response under assignment to HIV-1 vaccination (i.e., the treated potential outcome).\cite{Follman_2006} Since infant immune responses in the vaccine arm were more variable, adjustment for tetanus antibodies reduced the variance of the ATE substantially. In PACTG230, antibodies to gp41 were proxies for maternal antibody and the HIV-1-specific antibody response under hypothetical assignment to placebo. Adjustment for the gp41 antibody response reduced the variance of the ATE substantially when comparing each vaccine to placebo, but not when comparing vaccines head-to-head. Hence, our results suggest that the benefit of NCO adjustment may depend on whether the NCO is a proxy for the treated or treatment-free potential outcome, which causal contrast is of primary interest (e.g., placebo vs. vaccine or comparing several vaccines), and whether the vaccine or placebo arm is expected to contribute more variability to the ATE. 

\par While our work primarily focuses on small, early-phase trials measuring continuous immune response endpoints, improving efficiency of late-phase vaccine efficacy trials via adjustment for off-target infection endpoints warrants further exploration.\cite{etievant_increasing_2022} Our work has implications for study design of future vaccine trials in populations with prior exposure. We advocate for early-phase vaccine immunogenicity studies to measure a variety of auxiliary immune responses believed to be unaffected by the intervention -- such as baseline humoral immune responses, concurrent immune responses to vaccine-untargeted antigens, and immune responses to irrelevant vaccinations. Pooling data from placebo arms across several trials to identify valid NCOs and model relationships between auxiliary immune responses and immune responses of interest could hone the efficiency of future trials in the design and analysis stages.

%\backmatter
\bmsection*{Author contributions}

EA, HJ, BZ were involved in the conceptualization of the project. EA was responsible for methods development, numerical experiments, and initial drafting of manuscript. EA, HJ, GF, YF collaborated on the data application. All authors contributed to review and edits of manuscript.

\bmsection*{Acknowledgments}
This material is based upon work supported by the National Science Foundation Graduate Research Fellowship Program under Grant No. DGE-2140004. Any opinions, findings, conclusions, or recommendations expressed in this material are those of the authors and do not necessarily reflect the views of the National Science Foundation. The work was also supported by the National Institute of Allergy and Infectious Diseases (NIAID) under award number UM1AI068635.

The HIV Prevention Trials Network (HPTN) 027 study was funded by the US National Institutes of Health (NIH), initially through the HPTN and later through the International Maternal Pediatric Adolescent AIDS Clinical Trials (IMPAACT) group. The HPTN (U01AI46749) has been funded by the National Institute of Allergy and Infectious Diseases (NIAID), the Eunice Kennedy Shriver National Institute of Child Health and Human Development (NICHD), National Institute of Drug Abuse (NIDA), and National Institute of Mental Health (NIMH). The IMPAACT Group (U01AI068632) has been funded by NIAID, NICHD, and NIMH. The study product was provided for free by Sanofi-Pastuer.

Overall support for the International Maternal Pediatric Adolescent AIDS Clinical Trials Network (IMPAACT) was provided by the National Institute of Allergy and Infectious Diseases (NIAID) with co-funding from the Eunice Kennedy Shriver National Institute of Child Health and Human Development (NICHD) and the National Institute of Mental Health (NIMH), all components of the National Institutes of Health (NIH), under Award Numbers UM1AI068632 (IMPAACT LOC), UM1AI068616 (IMPAACT SDMC) and UM1AI106716 (IMPAACT LC), and by NICHD contract number HHSN275201800001I.  The content is solely the responsibility of the authors and does not necessarily represent the official views of the NIH.

\bmsection*{Financial disclosure}

None reported.

\bmsection*{Conflict of interest}

The authors declare no potential conflict of interests.

\bibliography{wileyNJD-AMA}

\newpage

\appendix

\section{Additional Simulation Results}

\subsection{Performance Under Violations of NCO Assumption}

As described in the main text of the paper, we explored the performance of NCO adjustment when the NCO assumption (Assumption 4) was violated to varying degrees. We assessed performance in terms of mean absolute bias relative to the plug-in estimator and coverage of nominal confidence intervals as a function of the magnitude of the NCO violation, sample size, and the predictiveness of the NCO. We varied the effect of treatment on the candidate NCO within $\beta_N \in \{0, 0.1, 0.25, 0.5, 0.75, 1, 1.5, 2\}$, encompassing cases where the NCO assumption is not violated and is violated to varying degrees. For simplicity, we focused on comparing the plug-in estimator to the NCO-adjusted estimator. We considered pairing the NCO-adjusted estimator with either a pretest of the sharp null hypothesis or an equivalence pretest at various equivalence thresholds $\epsilon > 0$. All pretests were conducted at level $\alpha=0.05$. If the sharp null hypothesis was rejected, the plug-in estimator would be returned, otherwise the NCO-adjusted estimator was returned. If the equivalence null hypothesis was rejected, the NCO-adjusted estimator was returned and otherwise, the plug-in estimator was returned. The effect on the primary outcome was set to $\beta=1$, and $\rho_{Y,X}=0.3$ and $\rho_{(Y,N|X)}$ was varied between $0.3$ and $0.8$. We focused on the identity link function as in Setting 1. Estimators were compared on the basis of bias and confidence interval coverage.

\begin{figure}
    \centering
    \includegraphics[width=0.9\linewidth]{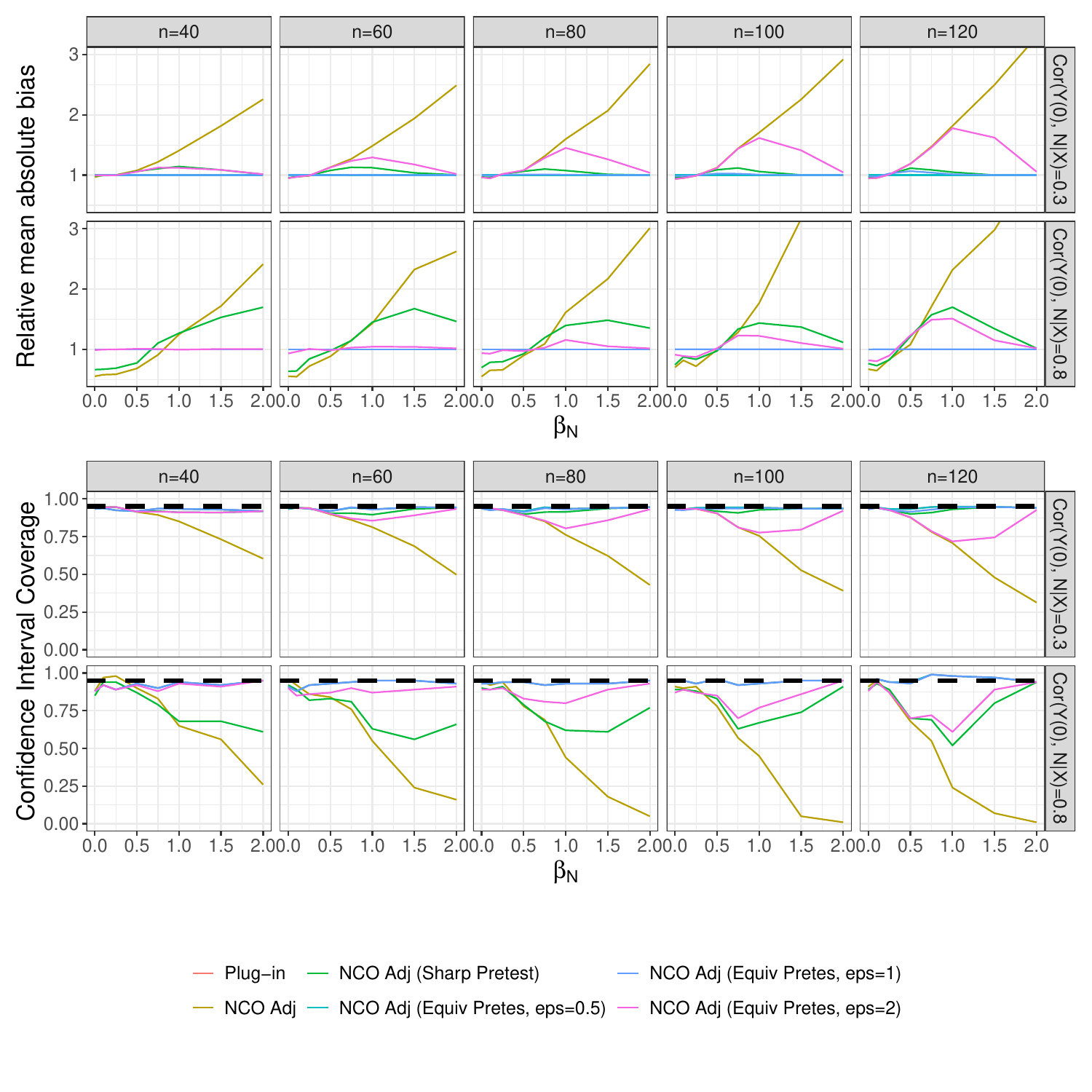}
    \caption{Numerical experiments illustrating the impact of NCO violations on the performance of NCO adjustment. Performance summarized using relative mean absolute bias and confidence interval coverage. Key parameters varied are sample size, correlation between $Y(0)$ and $N$ conditional on $X$, and the effect of treatment on the candidate NCO $\beta_N$.}
    \label{fig:viol}
\end{figure}

Results for the simulation are shown in Figure \ref{fig:viol}. Adjustment for the candidate NCO can lead to substantial relative bias as the effect of treatment on the candidate NCO increases. Imposing sharp null/equivalence pretests acts to mitigate the bias when the NCO assumption fails. However, most pretests are vulnerable to intermediate violations of the NCO assumption for small sample sizes. Under large violations of the NCO assumption, the pretest flags the violation and avoids bias. Under small violations, NCO adjustment proceeds as before but generates minimal bias. The performance of equivalence tests depends on the choice of equivalence threshold $\epsilon$. Small choices of equivalence threshold set very stringent thresholds to permit NCO adjustment, and therefore rarely adjust for the NCO and result in an estimator which mimics the plug-in. Larger equivalence thresholds only protect against violations of the NCO assumption that exceed the threshold and are more vulnerable to intermediate violations of the NCO assumption. The results for confidence interval coverage mirror those for the bias. NCO adjustment provides valid coverage when the NCO assumption is satisifed, but has very low coverage when the treatment has greater effect on the NCO candidate. Incorporating pretests guards against cases where NCO adjustment may be very harmful. As before, the performance of equivalence pretests depend on whether the violation of the NCO lies within or beyond the equivalence threshold.

\begin{figure}
    \centering
    \includegraphics[width=0.7\linewidth]{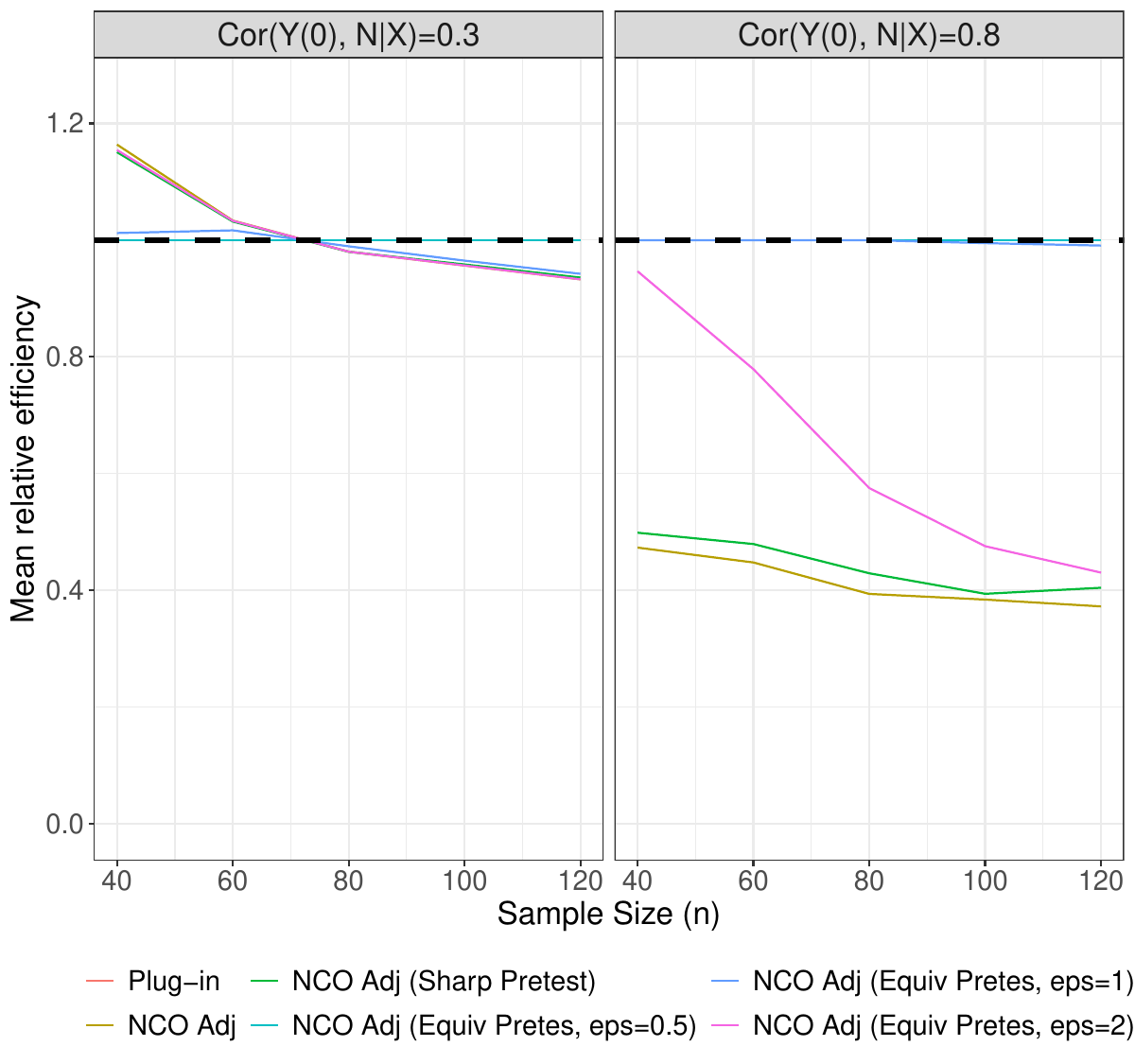}
    \caption{Performance of NCO adjustment (with and without pretests) when the NCO assumption is satisified ($\beta_N=0$). Key parameters varies are sample size ($n$) and correlation between $Y(0)$ and $N$ conditional on $X$.}
    \label{fig:viol2}
\end{figure}

Next, we compare the performance of the different estimators when the NCO assumption is satisifed to see whether there is any penalty paid for pretests in Figure \ref{fig:viol2}. In the case where the NCO is not very predictive, adjustment for the NCO leads to loss of precision in finite samples and a small precision gain in larger studies. When the NCO is highly prognostic, NCO adjustment leads to substantial gains in precision. In general, pretests reduce the benefit of NCO adjustment. Tests of the sharp causal null lead to a small precision loss relative to NCO adjustment outright, because the default assumption made by the sharp null is that the candidate NCO is valid. Most equivalence tests led to substantial precision losses, save for the equivalence test with the widest equivalence margin. As sample size increased, this equivalence pretest favored more frequent NCO adjustment, and average efficiency relative to the plug-in estimator decreased. 

\subsection{Results for Inference on SATE and Randomization Tests}

\begin{figure}
    \centering
    \includegraphics[width=\textwidth]{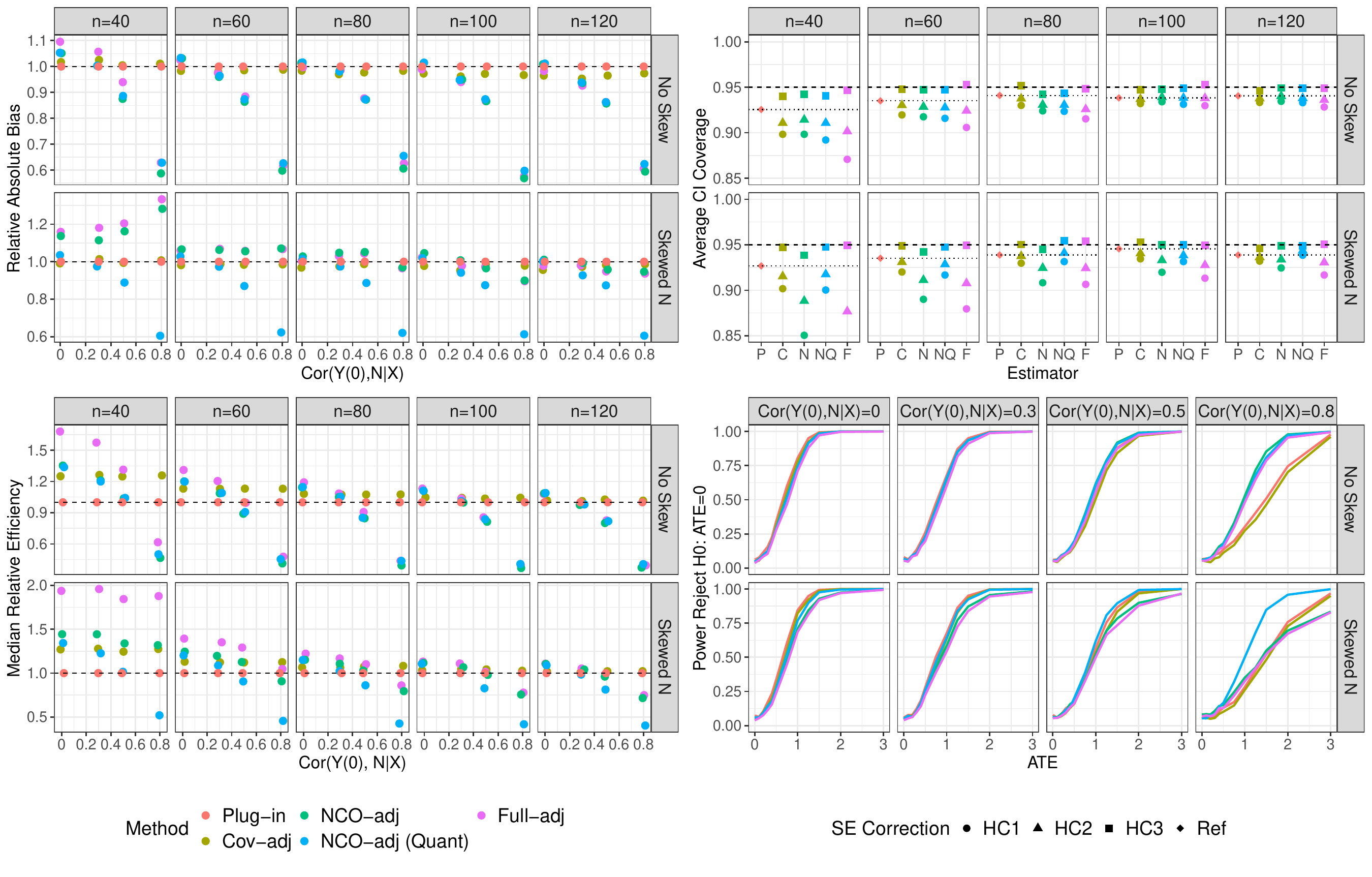}
    \caption{Top left: relative mean absolute bias of different estimators of the SATE as a function of sample size, data generating mechanism, and how prognostic the NCO is for the outcome of interest. Top right: average coverage of 95\% confidence intervals of different estimators of the SATE and finite sample standard error corrections as a function of sample size and data generating mechanism. Bottom left: median relative efficiency of estimators of the SATE as a function of sample size, data generating mechanism, and how prognostic the NCO is for the outcome of interest. Bottom right: power curves of Wald tests of H0: SATE=0 as a function of how prognostic the NCO is for the outcome of interest and the data generating mechanism.}
    \label{fig:SATEsim}
\end{figure}

In Figure \ref{fig:SATEsim}, we illustrate the performance of OLS-adjusted estimators of the sample average treatment effect (SATE) in numerical experiments in settings consistent with the main simulation described in the main text of the article. Similar to estimators of the population average treatment effect, we see that adjustment for the NCO can reduce finite-sample bias, increase precision, and increase power of Wald tests especially when the NCO becomes more prognostic for the primary outcome. When the NCO distribution is skewed, adjustment for the NCO on the quantile scale can mitigate the bias, efficiency loss, and power loss of estimators which adjust for the raw NCO. We observe that adjustment of the standard errors using the HC3 correction leads to improved coverage of confidence intervals in finite samples. Power to reject the weak null hypothesis $H_0^{\text{weak}}: \text{SATE}=0$ using robust t-statistics us shown in the bottom right panel for $n=60$. We observe that NCO-adjustment can lead to improvements in power to detect non-zero treatment effects when the NCO is more prognostic for the primary outcome.

\begin{figure}[h]
    \centering
    \includegraphics[width=0.75\textwidth]{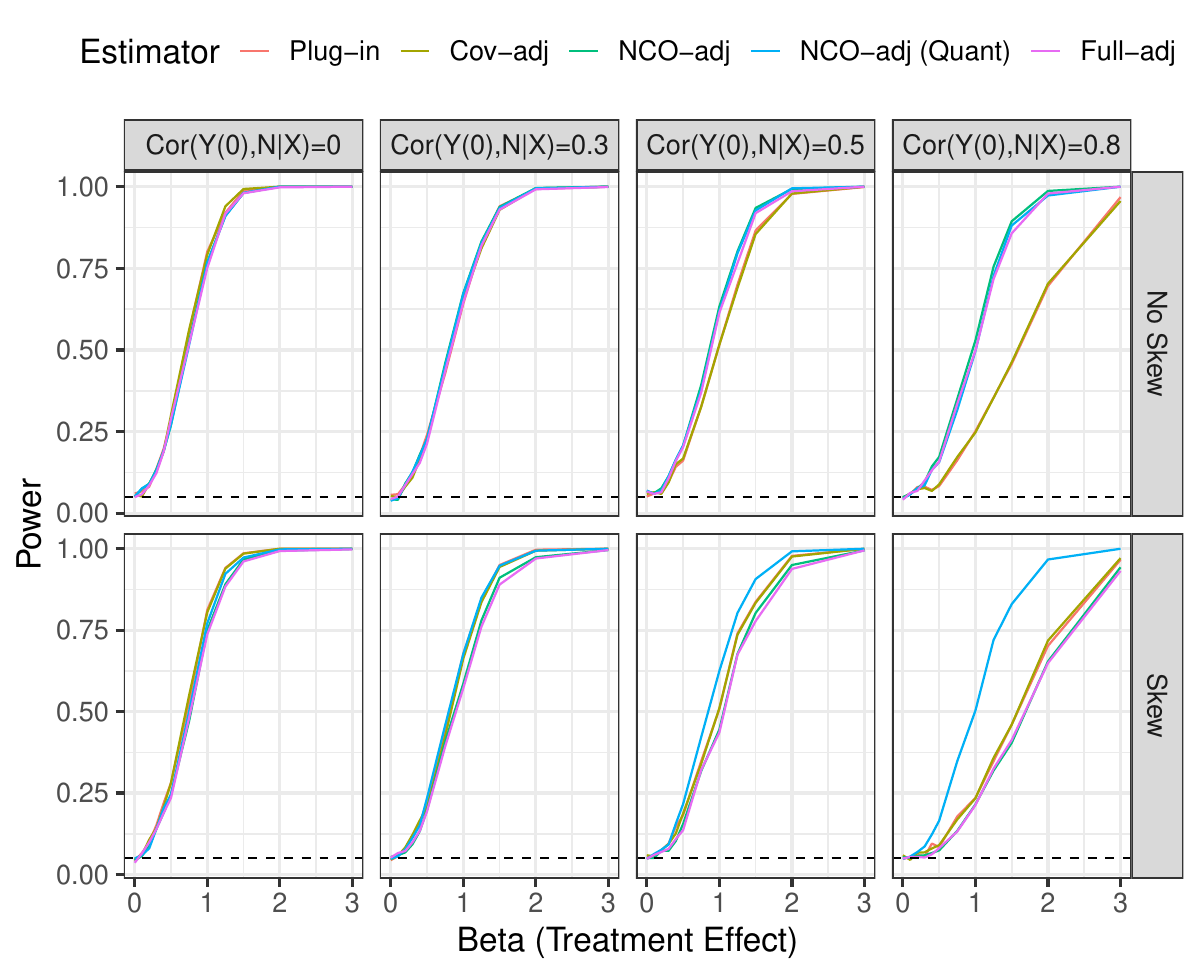}
    \caption{Power of randomization tests of $H_0^{\text{sharp}}: Y(1)=Y(0)$ across simulated datasets as a function of residual correlation between $Y$ and $N$, working model specification, and treatment effect.}
    \label{fig:supp1}
\end{figure}

In Figure \ref{fig:supp1}, we illustrate the power of unadjusted and adjusted randomization tests of $H_0^{\text{sharp}}: Y(1)=Y(0)$ across simulated datasets with sample size $n=60$ and randomization ratio $\pi=0.80$. We see that when $N$ and $Y$ exhibit very low correlations, the adjusted and unadjusted approaches have very similar power. However, when $N$ and $Y$ exhibit moderate to high correlation, adjustment for $N$ can lead to increases in statistical power to test the sharp causal null. In settings with the skewed predictor subject to detection limits, adjusting for the raw NCO on the quantile scale led to reduced power relative to covariate-adjusted and plug-in approaches. However, adjustment for the NCO on the quantile scale led to superior statistical power among all estimators. All estimators produced nominal Type I error rates under the null hypothesis of no treatment effect.

\subsection{Identities linking manipulable parameters to parameters of data generating process}

Recalling the data generating process in the main text. Note that
\begin{align*}
    \begin{pmatrix}
        Y \\
        N
    \end{pmatrix} \Big| X &\sim x\beta_1 + N(U\beta_2, \Sigma = \text{diag(1,2))} \\
    U|X &\sim N(0,1)
\end{align*}
A key parameter that we want to modulate between simulations is the partial correlation between $Y$ and $N$ conditional on $X$, defined as
\begin{align*}
    \rho_{Y,N|X} &= \text{Cov}(Y,N|X)/\sqrt{\text{Var}(Y|X)\text{Var(N|X)}}
\end{align*}
Using laws of total variance and covariance, we can express each of these unknowns according to the regression parameters
\begin{align*}
    \text{Var}(Y|X) &= \text{Var}(E[Y|X,U]|X) + E[\text{Var}(Y|X,U)|X] = \text{Var}(X\beta_1 + U\beta_2|X) + 1 = \beta_2^2 + 1\\
    \text{Var}(N|X) &= \text{Var}(E[N|X,U]|X) + E[\text{Var}(N|X,U)|X] = \text{Var}(X\beta_1 + U\beta_2|X) + 1 = \beta_2^2 + 1\\
    \text{Cov}(Y,N|X) &= \text{Cov}(E(Y|U,X),E(Y|U,X) \; |X) + \text{E}[\text{Cov}(Y,N|U,X) | X] = \text{Var}(X\beta_1 + U \beta_2 \; |X) + 0 = \beta_2^2
\end{align*}
Hence, 
\begin{align*}
    \implies \rho_{Y,N|X} &= \frac{\beta_2^2}{1+\beta_2^2} \\
    \beta_2 &= \sqrt{\frac{\rho_{Y,N|X}}{1-\rho_{Y,N|X}}}
\end{align*}

Next, we examine 
\begin{align*}
    (X,Y)^T &\sim N\left( (0, X\beta_1 + U\beta_2), \Sigma = \begin{pmatrix}
        1 & \rho_{X,Y} \\
        \rho_{X,Y} & 1
    \end{pmatrix}\right) \\
    X &\sim N(0,1)
\end{align*}
The unknown parameter can be expressed as 
\begin{align*}
    \rho_{X,Y} &= \text{Cov}(X,Y)/\sqrt{\text{Var}(Y)\text{Var(X)}}
\end{align*}
Note that we know that $\text{Var}(X)=1$. Now we can solve for the remaining components using laws of total variance/covariance.
\begin{align*}
    \text{Var}(Y) &= \text{Var}(E[Y|X,U]) + \text{E}[\text{Var}(Y|X,U)] = \text{Var}(X\beta_1 + U \beta_2) + 1 = \beta_1^2 + \beta_2^2 + 1 \\
    \text{Cov}(X,Y) &= \text{Cov}(E[X|X,U],E[Y|X,U]) + E[\text{Cov}(X,Y | X,U)] = \text{Cov}(X,\beta_1 X + \beta_2 U) + 0 = \beta_1
\end{align*}
Solving we obtain
\begin{align*}
    \rho_{X,Y} &= \beta_1/\sqrt{\beta_1^2 + \beta_2^2 + 1} \\
    \implies \beta_1^2 &=\rho_{X,Y}^2 (\beta_1^2 + \beta_2^2 + 1)   \\
    \implies \beta_1 &= \sqrt{\frac{\rho_{X,Y}^2 (\beta_2^2 + 1)}{1-\rho_{X,Y}^2}}
\end{align*}

\section{Graphical Causal Models Motivating NCO Adjustment}

\par In this section, we describe the causal models proposed in the DAG in the main text in more detail. Note that the causal diagrams in the DAG satisfy Assumptions 1-5. Assumption 3 is satisfied because no arrows point to treatment node $A$ and the baseline covariates $X$ are not caused by treatment $A$. Assumption 4 is satisfied by the absence of arrows from $A$ to $N$. Assumption 5 is satisifed by the absence of an arrow from $Y$ to $N$.
\par Suppose we focus on the causal model in the left panel of Figure the DAG. The causal diagram implies the following nonparametric structural equations model for the data.\citep{Pearl_1995}
\begin{align*}
    A &= f_{A}(\epsilon_A) \\
    U &= f_{U}(\epsilon_U) \\
    X &= f_{X}(U, \epsilon_X) \\
    N &= f_{N}(U, X, \epsilon_N) \\
    Y &= f_{Y}(U, X, N, A, \epsilon_Y)
\end{align*}
Where $(\epsilon_A, \epsilon_U, \epsilon_X, \epsilon_{Y}, \epsilon_N)$ are mutually independent random disturbances and the $f$ are unknown deterministic functions. From the NPSEM, we conceptualize $X$ and $N$ as error-prone proxies for an unmeasured variable $U$ that influences the primary outcome $Y$. In many cases, $X$ will be a limited set of characteristics (participant age, sex, and anthropometric measurements) that may very poorly capture the influence of $U$. The motivation for adjustment for an NCO is either (a) $N$ exerts a strong causal effect on $Y$ or (b) $N$ is a good proxy for the effect of $U$ on $Y$. 

Suppose we focus on the possibility outlined by (b). A strong assumption is that $N$ is a ``surrogate" for the effect of $U$ on $Y$ by the Prentice definition, meaning that $Y \perp U | N$. Under the surrogacy definition, we can revise the NPSEM for $Y$ as follows (with slight abuse of notation with respect to the deterministic generating function).
\begin{align*}
    Y = f_Y(X,N,A,\epsilon_Y)
\end{align*}
Under the surrogacy assumption, the observable set of variables $(X,N,A)$ comprises the oracle set of common causes of $Y$. Hence, $(X,N,A)$ contains the same predictive information for $Y$ as $(U,X,N,A)$; assuming working models are correctly specified, adjustment for $(X,N,A)$ will suffice in adjusting for $(U,X,N,A)$.

We provide a specific example to illustrate our point. Suppose we assume outcomes $Y$ and $N$ are continuous with an additive mean-zero error model for both $N$ and $Y$. Suppose the data generating models are partly linear in $U$.
\begin{align*}
    Y &= \beta U + f_{Y}(X, A) + \epsilon_Y \\
    N &= \beta U + f_{N}(X) + \epsilon_N
\end{align*}
Re-expressing $\beta U = N-f_N(X)-\epsilon_N$, we can write the data generating model for $Y$ as a partly linear model in $N$.
\begin{align*}
    Y &= \beta (N-f_N(X)-\epsilon_N) + f_{Y}(X, A) + \epsilon_Y \\
    \implies Y &= \beta N + f^*_{Y}(X, A) + \epsilon_Y^*
\end{align*}
Under a relatively simplistic model and the strong surrogacy assumption, we observe how the effect of an unmeasured variable $U$ on $Y$ can be indirectly obtained using the NCO $N$. If $\beta$ is large and the effect of $X$ is comparably smaller, then $Y$ will be approximately linear in $N$. In this case, adjustment for $N$ using OLS working models will render near-oracle efficiency (i.e., efficiency in an analysis where the full slate of common causes $U,X,N,A$ were known).

In the right panel of the DAG, we formalize the NPSEM associated with outcomes collected sequentially over trial follow-up and briefly comment on them. The oracle set of precision variables for the effect of treatment $A$ on the outcome vector $(Y_1, Y_2)$ is given by $(X, U_1, U_2)$. However, $(U_1, U_2)$ are an unmeasured time-varying process. Instead, we can adjust for $(X ,N_1, N_2)$, where $(N_1, N_2)$ can be considered as proxies for the unmeasured time-varying process $(U_1, U_2)$.
\begin{align*}
    A &= f_{A}(\epsilon_A) \\
    U_1 &= f_{U,1}(\epsilon_{U1}) \\
    U_2 &= f_{U,2}(U_1, \epsilon_{U2}) \\
    N_1 &= f_{Y,1}(U_1, \epsilon_{N1}) \\
    N_2 &= f_{Y,2}(U_2, U_1, N_1, \epsilon_{N2}) \\
    Y_1 &= f_{Y,1}(U_1, N_1, \epsilon_{Y1}) \\
    Y_2 &= f_{Y,2}(U_2, U_1, N_2, N_1, A, \epsilon_{Y2})
\end{align*}
Suppose we are interested in the effect of $A$ on $Y_2$. Suppose we make a similar surrogacy assumption as above, $Y_2 \perp (U_2, U_1) | (N_2,N_1)$. We then can see that the observable set of variables $(N_2, N_1, A)$ contains the same predictive information for $Y$ as the set of oracle variables $(U_2, U_1, N_2, N_1, A)$.

\section{Background on semiparametric theory \label{app1}}
\vspace*{12pt}
\par We provide a short primer on semiparametric theory which defines the useful terminology and concepts. In many classical statistical problems, we are interested in estimating a target quantity within a statistical model $M$ restricted to distributions indexed by a finite dimensional parameter $\theta$. Such models are referred to as parametric models. However, in many cases, we may not want to rely on a possibly-misspecified parametric model for inference for fear of incurring systematic bias that will not disappear as the sample size grows. Hence, we may restrict attention to distributions indexed by infinite dimensional parameters. A model $M$ indexed by infinite-dimensional parameter without any assumptions is referred to as the \textit{nonparametric model}. However, there are multitude of other restrictions that we could place on our model $M$ which may be motivated by scientific or experimental knowledge. A common set of assumptions we make, particularly in causal inference problems, are exclusion restrictions, or assumptions of independence between variables. For example, in a randomized trial measuring baseline covariates $X$, we can restrict focus to data generating laws satisfying the exclusion restriction $X \perp A$ which is satisfied by randomization. We refer to such models with some arbitrary restriction on the infinite-dimensional parameter as \textit{semiparametric models}.
\par In parametric models, maximum likelihood estimators are celebrated due to their \textit{efficiency} in large samples, meaning they asymptotically attain the Cramer-Rao lower bound, or the minimum possible variance an estimator can achieve under the parametric model. In semiparametric estimation problems, we are often interested in finding analogs to the parametric case. First, we are interested in what is the minimum possible variance that we can estimate a a target quantity across all data generating laws in the model. We refer to this quantity as the \textit{semiparametric efficiency bound}. Second, we are interested in identifying what estimator asymptotically achieves this minimum variance bound. We refer to this quantity as the \textit{semiparametric efficient estimator}.
\par We will restrict focus to regular asymptotically linear (RAL) estimators to avoid pathological estimators which display very unstable behavior uniformly over the parameter set. Parametric efficiency theory can be generalized to nonparametric and semiparametric statistical models by relying on the insight that the best possible variance attainable in an infinite-dimensional model $M$ should be \textit{at least as large} as the best possible variance attainable in \textit{any} parametric submodel $M_1 \subset M$. To obtain the tightest bound, we claim that the best possible variance in an infinite dimensional model is equal to the smallest achievable variance in the least favorable quadratic mean differentiable parametric submodel. Let $v_0^*(M)$ denote the variance of any RAL estimator in an infinite dimensional model. Let $\mathcal{H}(P_0)$ denote the collection of all smooth (quadratic mean differentiable) parametric submodels centered at $P_0$. We restrict focus to quadratic mean differentiable models as they will have score functions with mean zero and bounded variance. Let $M_h$ denote a particular choice of submodel. The generalized Cramer-Rao lower bound is as follows.
\begin{align*}
    v_0^*(M) &\geq \underset{h \in \mathcal{H}(P_0)}{\sup} v_0(M_h) = \underset{h \in \mathcal{H}(P_0)}{\sup} \frac{\left(\frac{\partial}{\partial \theta} \psi(P_{\theta, h}) \Big|_{\theta=0}\right)^2}{P_0 g_h^2}
\end{align*}
where $g_h$ is the score of the least-favorable parametric model through $h$. $P_0 g_h^2$ is the variance of the score function in the least favorable submodel. The numerator is a square of the pathwise derivative of the functional at $P_0$ evaluated along the submodel $h$. If $\psi$ is a pathwise differentiable functional, Riesz representation theorem guarantees the pathwise derivative is writable as an inner product of a gradient function $D(P_0)$ in the Hilbert space $L_0^2(P_0)$ and the score $g_h$ along the least favorable submodel. This yields a new and useful formulation of the efficiency bound.
\begin{align*}
    v_0*(M) \geq \underset{h \in \mathcal{H}(P_0)}{\sup} \frac{\left(P_0[D(P_0) g_h]\right)^2}{P_0 g_h^2}
\end{align*}
Recognizing that the variance bound depends on the set of parametric submodels $\mathcal{H}$ entirely through the score function $g_h$, we can rewrite the above bound in terms of the set of allowable scores $\mathcal{G}(P_0) := \{g_h : h \in \mathcal{H}(P_0)\}$, also known as the tangent space.
\begin{align*}
    v_0^*(M) \geq \underset{g \in \mathcal{G}(P_0)}{\sup} \frac{\left(P_0[D(P_0) g]\right)^2}{P_0 g^2}
\end{align*}
The above bound will have a closed form when the gradient $D(P_0) \in \mathcal{G}(P_0)$, meaning the gradient lies in the tangent space. In a nonparametric model, the set of allowable scores is all of $\mathcal{G}(P_0) = L_0^2(P_0)$, therefore, $D(P_0) \in \mathcal{G}(P_0)$ by default and the nonparametric efficiency bound is given by
\begin{align*}
    v_0^*(M) \geq P_0[D(P_0)^2]
\end{align*}
However, when we impose restrictions to our model $M$, we thin the collection of allowable scores in $\mathcal{G}(P_0)$ and the nonparametric gradient $D(P_0)$ is not guaranteed to lie in $\mathcal{G}(P_0)$ anymore. However, we can represent any gradient in the following manner.
\begin{align*}
    D(P_0) &:= D^*(P_0) + (D(P_0) - D^*(P_0))
\end{align*}
Where $D^*(P_0) \in \mathcal{G}(P_0)$ and $D(P_0) - D^*(P_0) \in \mathcal{G}(P_0)^\perp$ represents the orthogonal complement of the tangent space. We refer to $D^*(P_0)$ as the \textit{canonical gradient}. By substituting this expression into the generalized Cramer-Rao bound above
\begin{align*}
    v_0*(M) &\geq \underset{g \in \mathcal{G}(P_0)}{\sup} \frac{\left(P_0[(D^*(P_0) + (D(P_0) - D^*(P_0))) g]\right)^2}{P_0 g^2} \\
    &= P_0[D^*(P_0)^2]
\end{align*}
Which holds because $P_0((D(P_0) - D^*(P_0)) g) = 0$ because $D(P_0) - D^*(P_0) \in \mathcal{G}(P_0)^\perp$, meaning it is orthogonal to any score function $g \in \mathcal{G}(P_0)$.

This is a critical insight: the generalized Cramer-Rao lower bound in a semiparametric model depends on the \textit{orthogonal projection} of the gradient $D(P_0)$ onto the tangent space $\mathcal{G}(P_0)$. Hence, the recipe for obtaining the semiparametric efficiency bound is (a) identify a gradient in the nonparametric model $L^2(P_0)$ (which is guaranteed to be the unique gradient trivially by the uniqueness of a projection), (b) derive the form of the tangent space of allowable score functions in a semiparameteric model under some restrictions, and (c) compute the canonical gradient -- or the projection of the nonparametric gradient onto the tangent space comprised of scores compatible with the semiparametric model. The variance of the projection will represent the semiparametric efficiency bound.

Once these steps are completed, obtaining the efficient estimator in the semiparametric model is straightforward. A key insight in semiparametric estimation theory is that there is 1-1 correspondence between influence functions of RAL estimators and gradients of pathwise differentiable parameters. Hence, constructing a RAL estimator with influence function equal to the canonical gradient will produce the semiparametric efficient estimator of the target quantity.

\subsection{Proof of Proposition Proposition 1 \label{app1.1a}}

The following theorem uses the results from the previous subsection to derive the form of the efficiency bound and efficient influence function (EIF).

\textbf{Proof}: Let $A$ denote a binary treatment. Let $X$ denote a covariate, $N$ denote a valid NCO satisfying assumption 4, and $Y$ denote the outcome. Suppose Assumption 1-3 hold. hence, the following exclusion restriction holds: $A \perp (X,N,Y(0),Y(1))$. We define our statistical model $M$ as the set of all distributions for $(A,X,N,Y) \overset{iid}{\sim} P_0 \in M$ under the exclusion restriction above. Suppose our target parameter is the ATE, which is identified under Assumptions 1-3.
\begin{align*}
    \mathbb{E}[Y(1)-Y(0)] = \mathbb{E}[Y|A=1]-\mathbb{E}[Y|A=0]
\end{align*}
To derive the efficiency bound, we follow the recipe described in the previous subsection. Since the ATE is pathwise differentiable, there exists a correspondence between influence functions of RAL estimators and gradients. Indeed, one can show that the influence function for the nonparametric/plug-in estimator is 
\begin{align*}
    D(P_0)(a,y) = \frac{a(y-\mathbb{E}[Y|A=1])}{\pi} - \frac{(1-a)(y-\mathbb{E}[Y|A=0])}{1-\pi}
\end{align*}
We assume a known treatment probability $\pi$ such that the sample estimate $\pi = \pi$. One can also substitute a consistent estimator $\pi$ for $\pi$. The above influence function is also a gradient for our target parameter.

Step two in our recipe is to derive the form of the tangent space of $M$, i.e., the set of allowable score functions consistent with our model. We can decompose the into submodels for each constituent components.
\begin{align*}
    M &= M_{X,N} \otimes M_{A|X,N} \otimes M_{Y|X,N,A}
\end{align*}
Note that under our exclusion restriction $A \perp (X,N)$, $M_{A|X,N}=M_{A}$. This model for $A$ actually contains only a single density, $p(A|X,M) = \pi^{A}(1-\pi)^{1-A}$, where $\pi$ is known or consistently estimated. Since this component of the model consists of a single density, we can ignore this portion of the model when calculating the tangent space. We leave the other components of the model unspecified. Since we restricted focus to QMD models, we know the scores consist of all mean-zero, finite-variance functions. This implies they reside in the Hilbert space $L^2_0(P_0)$. Formally, the tangent space in model $M$ can be written as an orthogonal sum

\begin{align*}
    \mathcal{G} &= \mathcal{G}_{X,N} \oplus \mathcal{G}_{Y|X,N,A} \\
    \mathcal{G}_{X,N} &:= \{\text{all } g(X, N) \in L_{0, X,N}^2(P_0)\} \\
    \mathcal{G}_{Y|X,N,A} &:= \{\text{all } g(X,N,Y,A) \in L_{Y | X,N,A}^2(P_0)\}
\end{align*}

In the third step of our recipe, we must compute the projection of the gradient from step 1, $D(P_0)$, onto the tangent space $\mathcal{G}$ to obtain the canonical gradient. Recall that $D(P_0) = D^*(P_0) + D^\perp(P_0)$. Hence, any gradient is writable as the canonical gradient plus an additional term that lies in the orthogonal complement of the tangent space. As pointed out by \citep{tsiatis_covariate_2008}, any element of $\mathcal{G}$ can be written as
\begin{align*}
    Ah_1(X,N) + (1-A)h_0(X,N)
\end{align*}
for $h_1(X,N)$ and $h_0(X,N)$ both mean zero and finite variance. The moment restriction implied by randomization is
\begin{align*}
    \mathbb{E}[Ah_1(X,N) + (1-A)h_0(X,N)|X,N]&=0 \\
    \implies h_1(X,N) = -\frac{\pi}{(1-\pi)} h_0(X,N)
\end{align*}
Substituting into the original formula for an element of $\mathcal{G}$, we get that any element of $\mathcal{G}$ is writable as
\begin{align*}
    A \left(-\frac{\pi}{(1-\pi)} h_0(X,N)\right) + (1-A)h_0(X,N) \\
    \equiv -A\pi h_0(X,N) + (1-A)(1-\pi)h_0(X,N) \\
    \equiv (A-\pi) h_0(X,N)
\end{align*}
Hence, for a specific $h$, we can obtain the efficient influence function.
\begin{align*}
    \frac{a(y-\mathbb{E}[Y|A=1])}{\pi} - \frac{(1-a)(y-\mathbb{E}[Y|A=0])}{1-\pi} - (A-\pi) h(X,N)
\end{align*}

The specific $h$ we are looking for is the projection of the nonparametric gradient onto the orthogonal complement of the tangent space. Using some useful projection properties, we obtain that the projection takes the form
\begin{align*}
    \Pi(D(P_0)|\mathcal{G}^\perp) &= \frac{\mathbb{E}[Y|X,N,A=1]-\mathbb{E}[Y|A=1]}{\pi} - \frac{\mathbb{E}[Y|X,N,A=0]-\mathbb{E}[Y|A=0]}{1-\pi}
\end{align*}

Hence, the canonical gradient is given by
\begin{align*}
    D^*(P_0) &= \frac{a(y-\mathbb{E}[Y|A=1])}{\pi} - \frac{(1-a)(y-\mathbb{E}[Y|A=0])}{1-\pi} \\
    &- (A-\pi) \left(\frac{\mathbb{E}[Y|X,N,A=1]-\mathbb{E}[Y|A=1]}{\pi} - \frac{\mathbb{E}[Y|X,N,A=0]-\mathbb{E}[Y|A=0]}{1-\pi}\right) \\
    &= D(P_0) - (A-\pi)\left(\frac{\mathbb{E}[Y|X,N,A=1]}{\pi} - \frac{\mathbb{E}[Y|X,N,A=0]}{1-\pi}\right)
\end{align*}

The semiparametric efficiency bound is given by the variance of this gradient
\begin{align*}
    v_0^*(M_V) &\geq P_0[D^*(P_0)^2]
\end{align*}

By the 1-1 correspondence between gradients and influence functions of RAL estimators for pathwise differentiable parameters, we have that the RAL estimator that achieves the semiparametric efficiency bound, or the semiparametric efficient estimator of the ATE in a randomized trial with a NCO, takes the form
\begin{align*}
    \frac{1}{n} \sum_{i=1}^n D^*(P_0)(X_i, N_i, A_i, Y_i) = \bar{Y}(1) - \bar{Y}(0) - \frac{1}{n}\sum_{i=1}^n (A_i-\pi)\left(\frac{\mathbb{E}(Y|X,N,A=1)}{\pi} - \frac{\mathbb{E}(Y|X,N,A=0)}{1-\pi}\right)
\end{align*}

We argue that the estimator which adjusts for $(X, N)$ offers guaranteed efficiency gain over estimators which adjust for $X$ alone. We argue this point using the following result: let $M_1$ be a statistical model satisfying $A \perp (X,N)$ and $M_2$ be a statistical model satisfying $A \perp X$. Clearly, $M_1 \subseteq M_2$. By properties of gradients in nested models, every gradient in $M_2$ is also a gradient in $M_1$. Hence, the canonical gradient (i.e., efficient influence function) in $M_2$ is a gradient in $M_1$, but may not necessarily be the canonical gradient/efficient influence function. Hence, the variance lower bound achievable in $M_1$ is at most variance lower bound achievable in $M_2$. \qed

\subsection{Proof of Corollary 1}

Consider the oracle augmented data generating mechanism which includes measurement of the latent variable $U$ in the model where $(X,U,N,Y(1),Y(0)) \perp A$.
\begin{align*}
    (A_i, X_i, U_i, N_i, Y_i) \overset{i.i.d}{\sim} P_0 \in M^*
\end{align*}
Following the same logic as in the Proof of Proposition 1, the canonical gradient for the ATE in this expanded model where $(X,U,N)$ is measured is given by
\begin{align*}
    D_1^*(P_0) &= \frac{a(y-\mathbb{E}[Y|A=1])}{\pi} - \frac{(1-a)(y-\mathbb{E}[Y|A=0])}{1-\pi} \\
    &- (A-\pi) \left(\frac{\mathbb{E}[Y|X,U,N,A=1]-\mathbb{E}[Y|A=1]}{\pi} - \frac{\mathbb{E}[Y|X,U,N,A=0]-\mathbb{E}[Y|A=0]}{1-\pi}\right) \\
    &= D(P_0) - (A-\pi)\left(\frac{\mathbb{E}[Y|X,U,N,A=1]}{\pi} - \frac{\mathbb{E}[Y|X,U,N,A=0]}{1-\pi}\right)
\end{align*}
Where $D(P_0)$ is the gradient in the nonparametric (unrestricted model).

Suppose we make the additional assumption that $Y$ is \textit{mean independent} of $X,U$ conditional on $N, A$, or formally,
\begin{align*}
    \mathbb{E}[Y|X,U,N,A=a] = \mathbb{E}[Y|N,A=a]
\end{align*}
This assumption is related to the idea of surrogacy, as $N$ is taken as a surrogate for the effect of $(X,U)$ on $Y$. The canonical gradient in the randomized trial satisfying $A \perp (X,U,N)$ under the mean independence assumption is given by
\begin{align*}
    D_2^*(P_0) &= D(P_0) - (A-\pi)\left(\frac{\mathbb{E}[Y|N,A=1]}{\pi} - \frac{\mathbb{E}[Y|N,A=0]}{1-\pi}\right)
\end{align*}
The semiparametric efficiency bound, or the minimal variance in the semiparametric model of laws satisfying (i) $A \perp (X,U,N)$ and (ii) mean independence of $Y$ and $X,U$ conditional on $N,A$ is equal to the second moment of the canonical gradient, $P_0[D_2^*(P_0)^2]$. Recognizing the 1-1 correspondence between gradients and influence functions of RAL estimators for pathwise differentiable parameters, we have that the RAL estimator with the canonical gradient as its influence function achieves the semiparametric efficiency bound. The semiparametric efficient estimator in the augmented data model under mean independence is given by the following.
\begin{align*}
    \frac{1}{n} \sum_{i=1}^n D_2^*(P_0)(N_i, A_i, Y_i) = \frac{1}{n} \sum_{i=1}^n \frac{A_iY_i}{\pi} - \frac{(1-A_i)Y_i}{1-\pi} - (A_i-\pi)\left(\frac{\mathbb{E}(Y|N,A=1)}{\pi} - \frac{\mathbb{E}(Y|N,A=0)}{1-\pi}\right)
\end{align*}

Furthermore, suppose that $h_0$ and $h_1$ are substituted for the unknown regression functions in the formula for the influence function. By virtue of belonging in the class of AIPW estimators, the estimator will be consistent and asymptotically normal regardless of choices of the working models. Suppose we use linear working models fit using OLS for the full (oracle) data unit and a reduced data unit that does not measured $(X,U)$. Let the unknown parameters below denote the limits in probability of the parameters estimates obtained by OLS (note that estimates need not converge to the true parameters, merely some fixed limit).
\begin{align*}
    h_{a}(X,U,N) &= \beta_0 + \beta_1 X + \beta_2 U + \beta_3 N \\
    h^*_{a}(N) &= \beta_0^* + \beta_3^* N
\end{align*}
We rely on the following idea that if two variables are independent, they are also uncorrelated. By assumption, $Y \perp (U,X) | (N,A)$, implying that $\beta_1 = \beta_2 = 0$ in the first working model. It follows that the two working models specified above are asymptotically equivalent, as the probability limits of the parameter estimates satisfy $\beta_0^* = \beta_0$ and $\beta_3 = \beta_3^*$ because $\beta_1=\beta_2=0$. Hence, OLS working models based on the oracle data unit $\{X,U,N\}$ are asymptotically equivalent to OLS working models with $N$ as the sole predictor, implying that $\hat{\psi}_{\text{NCO-AIPW}}$ is asymptotically equivalent to $\hat{\psi}_{\text{Oracle-AIPW}}$, the AIPW OLS-assisted estimator which adjusts for the oracle data unit.

As review, when using OLS regression models, $\hat{\psi}_{\text{Cov-AIPW}}$ is guaranteed to be more efficient than $\hat{\psi}_{\text{Plug-in}}$.\cite{tsiatis_covariate_2008} Adjusting for more explanatory variables cannot harm large-sample efficiency. Hence, $\hat{\psi}_{\text{Oracle-AIPW}}$ is guaranteed to be at least as efficient as $\hat{\psi}_{\text{Cov-AIPW}}$. As described above, because the probability limits of the working models $h_{a}(X,U,N)$ are equivalent to the probability limits of the working models $h^*_{a}(N)$, $\hat{\psi}_{\text{NCO-AIPW}}$ is asymptotically equivalent to $\hat{\psi}_{\text{Oracle-AIPW}}$. This implies that $\hat{\psi}_{\text{NCO-AIPW}}$ offers guaranteed efficiency gain relative to $\hat{\psi}_{\text{Cov-AIPW}}$ and no efficiency loss relative to $\hat{\psi}_{\text{Oracle-AIPW}}$ respectively, even when the working models are misspecified.\qed

\section{Sensitivity Analysis}

In the main portion of this paper, we propose using randomization tests of the sharp null or an equivalence test to identify violations of the NCO assumption. Below, we describe a sensitivity analysis based on linear structural models which could be used to investigate the impacts of assumption violations on the ATE.

\subsection{Sensitivity Analysis Using Linear Models}

We can pursue a sensitivity analysis to the NCO assumption based on linear models as described by Rosenbaum\citep{Rosenbaum_1984}. The bias of an estimator which adjusts for a variable affected by treatment in a randomized experiment is the result of marginalizing over the factual distribution of $N$ rather than the proper counterfactual distributions of $N(1), N(0)$:
\begin{align*}
    \text{Bias} = &\mathbb{E}_{N}[\mathbb{E}_0[Y(1) | N = n] - \mathbb{E}_0[Y(0) | N = n]] - \left[\mathbb{E}_{N(1)}[\mathbb{E}_0[Y(1) | N = n] - \mathbb{E}_{N(0)}[\mathbb{E}_0[Y(0) | N = n]]\right] \\
    &\equiv (\mathbb{E}_{N}-\mathbb{E}_{N(1)})\{\mathbb{E}_0[Y(1) | N = n]\} - (\mathbb{E}_{N}-\mathbb{E}_{N(0)})\{\mathbb{E}_0[Y(0) | N = n]\}
\end{align*}
Where $\mathbb{E}_{N}[\cdot]$ refers to expectation taken over the observed distribution of $N$ and $\mathbb{E}_{N(a)}[\cdot]$ refers to expectation taken over the distribution of counterfactuals $N(a)$. When Assumption 4 holds and $N=N(1)=N(0)$, the bias term is equal to zero exactly.

Let $P_{N(a)}$ denote the measure of the counterfactual post-baseline outcome and $P_{N} := \alpha P_{N(1)} + \beta P_{N(0)}$ denote the measure of the observed distribution of $N$, which is necessary a mixture of the two counterfactual distributions such that $\alpha, \beta \in [0,1]$ and $\alpha + \beta = 1$. We can rewrite the expectations in empirical process form.

\begin{align*}
    \text{Bias} = \int \mathbb{E}_0[Y(1)|N=n] d((\alpha-1) P_{N(1)} + \beta P_{N(0)}) - \int \mathbb{E}_0[Y(0)|N=n] d((\beta-1) P_{N(0)} + \alpha P_{N(1)}) 
\end{align*}

Suppose we assume the following parallel linear structural nested models for the outcomes.
\begin{align*}
    \mathbb{E}_0[Y(1)|N=n] &= \alpha_1 + \gamma N \\
    \mathbb{E}_0[Y(0)|N=n] &= \alpha_0 + \gamma N
\end{align*}

The bias can be expressed as

\begin{align*}
    \text{Bias} &= \int \alpha_1 + \gamma N \; d((\alpha-1) F_{N(1)} + \beta F_{N(0)}) - \int \alpha_0 + \gamma N \; d((\beta-1) F_{N(0)} + \alpha F_{N(1)}) \\
    &= \gamma \int N d (F_{N(0)} - F_{N(1)}) = - \gamma \left(\mathbb{E}_0[N(1)-N(0)]\right)
\end{align*}

The bias of the adjusted estimator is linear in the average treatment effect on the post-baseline outcome. One can estimate $\gamma$ by fitting an OLS estimators with main effects for $(A,N)$ and no interaction, and using the estimated regression coefficient on $N$. Then, sensitivity analysis can be performed by choosing plausible values for the effect of treatment on the auxiliary outcomes, $\mathbb{E}_0[N(1)-N(0)]$, and subtracting the bias from the estimated treatment effect.

\end{document}